\documentclass[english,pra,aps,twocolumn,letterpaper,superscriptaddress]{revtex4-2}

\usepackage[T1]{fontenc}

\usepackage{graphicx}
\usepackage{epstopdf}
\usepackage{amssymb}
\usepackage{amsmath}
\usepackage[urlcolor=blue,hyperindex,colorlinks,bookmarks=true,linkcolor=black,citecolor=black]{hyperref}
\usepackage[normalem]{ulem}
\usepackage[abs]{overpic}
\usepackage{color}
\usepackage{rotating}
\usepackage{subfigure}
\usepackage{physics}

\usepackage[english]{babel}
\usepackage{blindtext,tikz}
\usetikzlibrary{calc}

\newcommand{\be}{\begin{equation}}
\newcommand{\ee}{\end{equation}}
\newcommand{\bea}{\begin{align}}
\newcommand{\eea}{\end{align}}
\def\iden{\mathbb{I}}
\newcommand{\kk}{\kappa}

\newcommand{\al}{\alpha}
\newcommand{\ww}{\omega}

\newcommand{\hrho}{\hat{\rho}}

\begin{document}

\title{Stabilizing two-qubit entanglement by mimicking a squeezed environment}

\author{L.~C.~G.~Govia}
\email[Electronic address: ]{lcggovia@gmail.com}
\affiliation{Quantum Engineering and Computing, Raytheon BBN Technologies, 10 Moulton St., Cambridge, MA, USA.}
\affiliation{Pritzker School of Molecular Engineering, University of Chicago, Chicago, IL, USA.}
\author{A.~Lingenfelter}
\affiliation{Pritzker School of Molecular Engineering, University of Chicago, Chicago, IL, USA.}
\author{A.~A.~Clerk}
\affiliation{Pritzker School of Molecular Engineering, University of Chicago, Chicago, IL, USA.}

\begin{abstract}

  It is well known that qubits immersed in a squeezed vacuum environment exhibit many exotic phenomena, including dissipative entanglement stabilization. Here, we show that these effects only require interference between excitation and decay processes, and can be faithfully mimicked without non-classical light using simple classical temporal modulation.  We present schemes that harnesses this idea to stabilize entanglement between two remote qubits coupled via a transmission line or waveguide, where either the qubit-waveguide coupling is modulated, or the qubits are directly driven.  We analyze the resilience of these approaches against various imperfections, and also characterize the trade-off between the speed and quality of entanglement stabilization.  Our protocols are compatible with state of the art cavity QED systems.     

\end{abstract}

\maketitle

%%%%%%%%%%%%%%%%%%%%%%%%%%%%%%%%%%%%%%%%%
%%%%%%%%%%%%%%%%%%%%%%%%%%%%%%%%%%%%%%%%%
%%%%%%%%%%%%%%%%%%%%%%%%%%%%%%%%%%%%%%%%%

%%%%%%%%%%%%%%%%%%%%%%%%%%%%%%%%%%%%%%%%%
\section{Introduction}

Squeezed states of electromagnetic radiation \cite{QuantumNoise,Walls:1983aa}, have
long been of interest, with applications ranging from quantum metrology and sensing \cite{Giovannetti1330,Caves:1980aa,Braginsky:1992kq,Aasi2013,Iwasawa:2013aa,Peano:2015aa,Taylor:2013aa}, to qubit readout \cite{Barzanjeh2014,Didier2015,Didier:2015aa,Govia:2017aa,Eddins:2018aa} and enhanced quantum gates \cite{Puri:2016aa,Royer2017}. 
Driving systems with squeezed vacuum noise can also lead to interesting dissipative physics.  A single qubit driven by broadband squeezed vacuum noise is described by the master equation \cite{PhysRevLett.56.1917}
\begin{align}
    \dot\hrho_q = \gamma\mathcal{D}\left[\cosh(r)\hat{\sigma}_- + \sinh(r)\hat{\sigma}_+ \right]\hrho_q, \label{eqn:qSMS}
\end{align}
where $\hat{\sigma}_\pm$ are standard qubit raising/lowering operators, $\mathcal{D}[x]\hrho = x\rho x^\dagger - \left\{x^\dagger x,\hrho\right\}/2$ is the usual Lindblad dissipator, $\gamma$ is the qubit-bath coupling rate, and $r$ is the standard squeezing parameter. 
While the steady state of Eq.~(\ref{eqn:qSMS}) is mixed, it generates non-trivial dynamics, with consequences including the existence of two distinct transverse relaxation times \cite{PhysRevLett.56.1917,Murch:2013aa} and modifications of the Mollow triplet fluorescence spectrum \cite{Carmichael:1987aa,Carmichael:1987ab,Dalton:1999aa,Toyli:2016aa}.

The dissipative dynamics becomes even richer when two qubits are coupled to a broadband two-mode squeezed vacuum (TMSV) environment, as the dynamics can now prepare and stabilize remote qubit entanglement \cite{Kraus:2004aa}. The evolution is described by
\begin{align}
  \nonumber\dot\hrho_q &= \gamma_1\mathcal{D}\left[\cosh(r)\hat{\sigma}_-^{(1)} + \sinh(r)\hat{\sigma}_+^{(2)}\right]\hrho_q \\ &+ \gamma_2\mathcal{D}\left[\sinh(r)\hat{\sigma}_+^{(1)} + \cosh(r)\hat{\sigma}_-^{(2)}\right]\hrho_q, \label{eqn:qTMS}
\end{align}
where $\gamma_{1/2}$ are the qubit-bath coupling rates, and $r$ is the squeezing parameter of the TMSV. The photon number pairing correlations of the TMSV environment are inherited by the qubits, leading to a pure entangled steady state having the form
\begin{align}
\ket{\Psi}_q = \frac{1}{\sqrt{\cosh(2r)}}\left(\cosh(r)\ket{00} - \sinh(r)\ket{11}\right). \label{eqn:EntQ}
\end{align}

%%%%%%%%%%%
% Fig. 1
%%%%%%%%%%%
\begin{figure}[t]
  \includegraphics[width=\columnwidth]{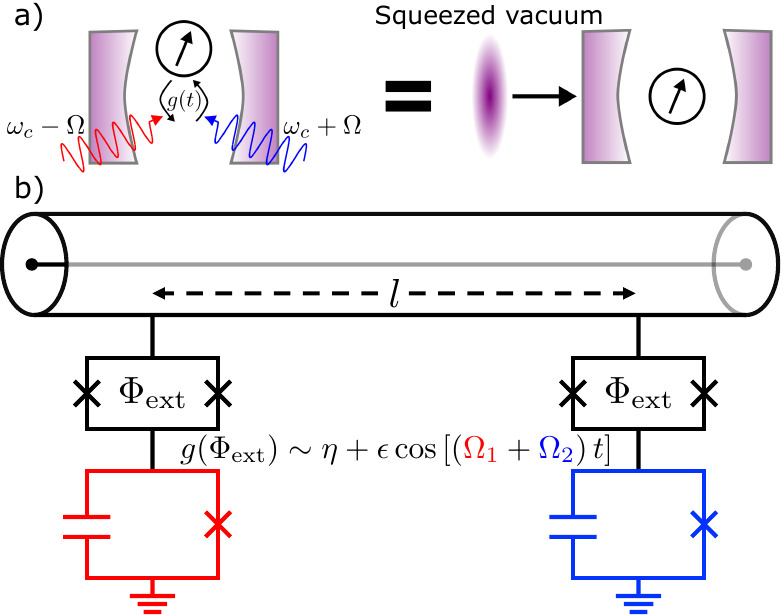}
  \caption{{\bf Synthetic squeezing}. {\bf a)} The qubit dynamics produced by biharmonic modulation of the qubit-cavity coupling are equivalent to that for injected squeezed vacuum resonant with the cavity. {\bf b)} Circuit QED schematic of two-qubit remote entanglement generation via synthetic squeezing. SQUID-modulated coupling between the qubits and a transmission line implements the dynamics of Eq.~\eqref{eqn:qTMS}.}
  \label{fig:setup}
\end{figure}
%%%%%%%%%%%

The dissipative physics of both Eqs.~(\ref{eqn:qSMS}) and (\ref{eqn:qTMS}) are often taken to be hallmarks of the interaction of matter with non-classical radiation.  Unfortunately, realizing these ``dissipative squeezing'' ideas experimentally is severely limited by the difficulty in generating, transporting and injecting squeezed states with high efficiency. 
In this work, we point out an underappreciated fact:  the non-trivial dissipative dynamics of both Eqs.~(\ref{eqn:qSMS}) and (\ref{eqn:qTMS}) can be realized {\it without the use of any non-classical radiation}, but instead by directly engineering interference between relevant qubit excitation and relaxation pathways.  As we show in detail, these can be achieved in a variety of systems where the the qubit frequencies and/or couplings to a vacuum reservoir can be temporally modulated (depicted in Fig.~\ref{fig:setup}a)).  We term these schemes ``synthetic squeezing'', and show that they provide a powerful route towards entanglement stabilization.

Our work discusses several methods for implementing synthetic squeezing.  This is of more than just academic interest: it also provides a potentially powerful new method for preparing and stabilizing entangled states between remote qubits using only classical drives.  We discuss two versions of this idea:  a first method that uses time-modulated couplings between qubits and a photonic link (i.e.~a waveguide or transmission line structure), and a second method that only uses local Rabi drives on each qubit. Schemes similar to our first approach were studied in Refs. \cite{ma_coupling-modulation-mediated_2021,ma_stabilizing_2019} but the equivalence to driving with squeezed light was not discussed. The second approach both generalizes and provides intuition for a protocol first introduced in Ref.~\cite{Motzoi:2016aa}. Including the impact of intrinsic thermal dissipation, we show that two-qubit entanglement stabilization with concurrence above $90\%$ is achievable in contemporary circuit QED devices.

While relatively unexplored in the few qubit case, we note that the general idea of mimicking squeezed reservoirs with parametric or Raman processes has been studied in the context of stabilizing bosonic states.  This includes schemes for producing bosonic squeezed states in 
trapped ions \cite{Cirac1993},
optomechanics \cite{Kronwald2013} and circuit QED \cite{Didier:2014aa}, and bosonic optomechanical entanglement \cite{Woolley:2013aa,Woolley:2014aa}.
Related master equations have also been studied in the setting of spin ensembles \cite{Parkins:2006aa,Dalla-Torre:2013aa}, though the specific equivalence to driving with squeezed light was not discussed.  

The remainder of this paper is organized as follows. In Secs.~\ref{sec:1qubit} and \ref{sec:2qubit} we present our synthetic squeezed dissipation schemes for both the single and two qubit cases. In Sec.~\ref{sec:SSE}, we study the approach to steady-state of the two-qubit scheme, its connections to large-spin entanglement stabilization, and the robustness of the steady-state entanglement to thermal dissipation. Finally, in Sec.~\ref{sec:conc} we present our concluding remarks.

\section{Single Qubit Synthetic Squeezed Dissipation Via Coupling Modulation}
\label{sec:1qubit}

%%%%%%%%%%%
% Fig. 2
%%%%%%%%%%%
\begin{figure}[t]
  \includegraphics[width=\columnwidth]{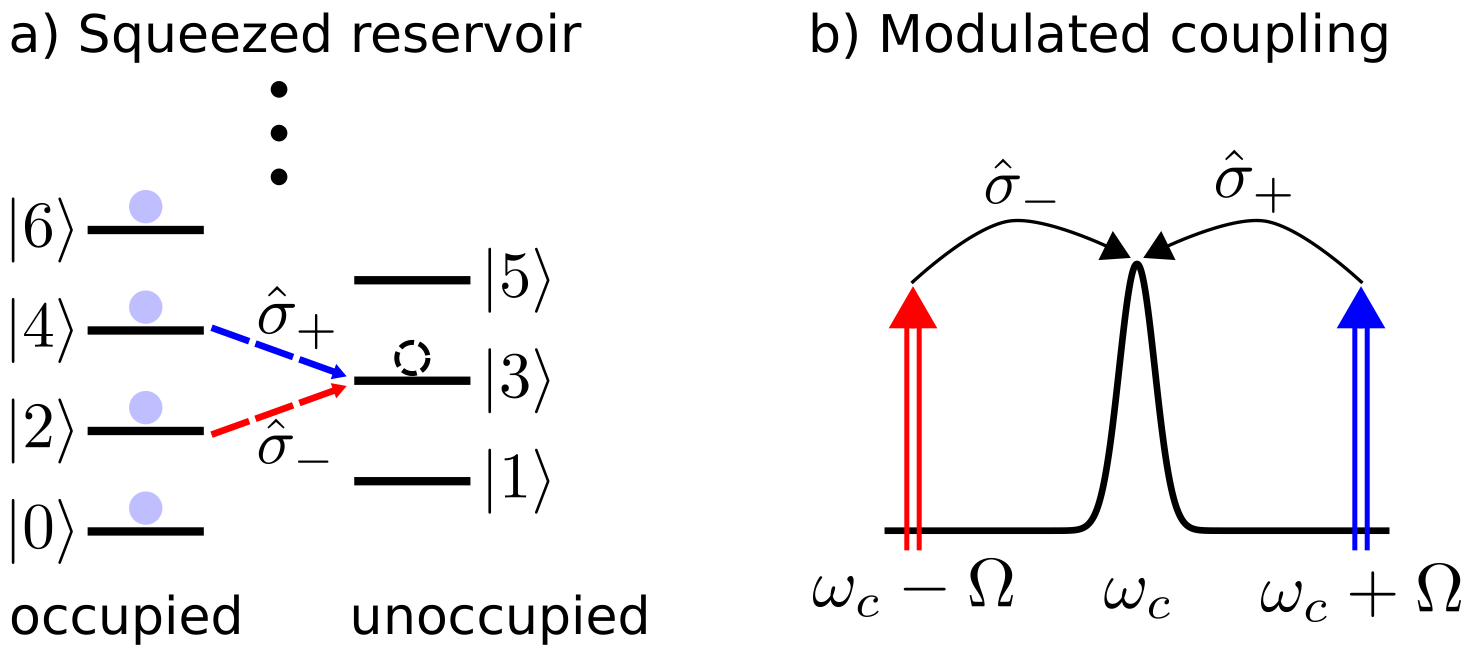}
  \caption{{\bf Pathways to the dissipative dynamics of Eq.~\eqref{eqn:qSMS}}. {\bf a)} Interference in the qubit excitation and decay channels when subject to a squeezed vacuum environment. {\bf b)} The blue and red sideband processes driven by coupling modulation at $\omega_c \pm \Omega$ interfere such that both qubit decay (red sideband) and excitation (blue sideband) result in an emitted cavity photon at frequency $\omega_c$.}
  \label{fig:int}
\end{figure}
%%%%%%%%%%%

We begin by describing our scheme for synthetic dissipative squeezing of a single qubit, i.e.~Eq.~\eqref{eqn:qSMS}.  The starting point is a cavity-qubit system with a time dependent coupling, described by the Hamiltonian:
\begin{align}
  \hat{H} = \omega_c\hat{a}^\dagger\hat{a} + \frac{\Omega}{2}\hat{\sigma}_z + g(t)\left(\hat{a}+\hat{a}^\dagger\right)\hat{\sigma}_x.
\end{align}
This is the usual Rabi Hamiltonian describing a cavity of frequency $\omega_c$ coupled to a qubit of frequency $\Omega$, with a time-dependent coupling $g(t)$. We consider phase-locked double-sideband cosine modulation of the coupling at frequencies $\omega_c \pm \Omega$, described by
\begin{align}
g(t) = \bar{g}\left(\al_+e^{-i(\omega_c + \Omega)t} + \al_-e^{-i(\omega_c - \Omega)t}\right) + h.c., \label{eqn:coup}
\end{align}
with $\al_{\rm \pm}$ the complex-valued amplitudes of the modulation. In circuit QED systems, this parametrically modulated coupling has been directly demonstrated using a superconducting quantum interference device (SQUID) \cite{Chen:2014aa,McKay:2016aa,Lu:2017aa,Zhong:aa,ClelandPRL2020,Noh2021,Brown2021}, and alternatively implemented using a dispersive qubit-cavity coupling, a qubit Rabi drive, and classical cavity displacements \cite{Hacohen-Gourgy:2016aa,Eddins:2018aa}.

After going to the interaction frame for the qubit and cavity, we make a rotating-wave approximation (RWA) to drop all time-dependent terms that oscillate faster than $\Delta = \omega_c-\Omega$, and obtain the Hamiltonian
\begin{align}
\hat{H}' &= \bar{g}\hat{a}^\dagger\left(\al_+\hat{\sigma}_{+} + \al_-\hat{\sigma}_{-}\right) +h.c., \label{eq:HInt}
\end{align}
where we require that $\abs{\al_{\pm}} \ll 2\Delta/\bar{g}$ for the RWA. Assuming that the cavity decays at a rate $\kk \gg \abs{\al_\pm} \bar{g}$ into a zero temperature environment, adiabatic elimination of the cavity yields a qubit-only master equation
\begin{align}
  \dot\hrho_q = \frac{4\bar{g}^2}{\kk}\mathcal{D}\left[\al_+\hat{\sigma}_+ + \al_-\hat{\sigma}_-\right]\hrho_q. \label{eqn:SynqSMS}
\end{align}
With the associations $\al_+ = \alpha\sinh(r)$ and $\al_- = \alpha\cosh(r)$ we obtain Eq.~\eqref{eqn:qSMS} with decay rate $\gamma =4\bar{g}^2\alpha^2/\kk$ and squeezing parameter $r$ set by $\abs{\al_+/\al_-} = \tanh(r)$.  We thus have our first (and simplest) example of how classical time-modulation can be use to simulate the dissipative effects of a squeezed environment.
We stress that this protocol does not require any input squeezed light, nor does it generate any bosonic squeezing in the cavity mode.
Note that modulation of tuneable couplings to control qubit cooling and heating processes has been studied previously \cite{Lu:2017aa,Huang:2018aa}, but two-tone modulation and the connection to squeezing was not discussed.

It is worth emphasizing the heuristic reasons why interfering sidebands mimic the dissipative effects of squeezed vacuum.  For a squeezed vacuum reservoir, the interference of qubit excitation and relaxation processes encoded in Eq.~(\ref{eqn:qSMS}) is a direct consequence of the photon-number pairing in the squeezed state.  To be explicit, if the reservoir ends up in a state with an odd number of photons $2M+1$, this could have resulted from a qubit excitation process starting with $2M+2$ reservoir photons, {\it or} from a qubit relaxation process starting from $2M$ reservoir photons (see
Fig.~\ref{fig:int}a)).  In our synthetic scheme, an identical interference is achieved using only classical radiation, i.e.~the coupling modulation.  The interference is now between the blue sideband Raman process (driven by coupling modulation at frequency $\ww_c+\Omega$) with the red sideband Raman process (driven by the modulation at frequency $\ww_c-\Omega$).  Both processes result in the generation of a cavity photon,  as depicted graphically in Fig.~\ref{fig:int}b).    

%%%%%%%%%%%%%%%%%%%%%%%%%%%%%
%%%%%%%%%%%%%%%%%%%%%%%%%%%%%

\section{Two-Qubit Synthetic Squeezed Dissipation via Coupling Modulation}
\label{sec:2qubit}

%%%%%%%%
%Fig. 3
%%%%%%%%
\begin{figure}
  \includegraphics[width=\columnwidth]{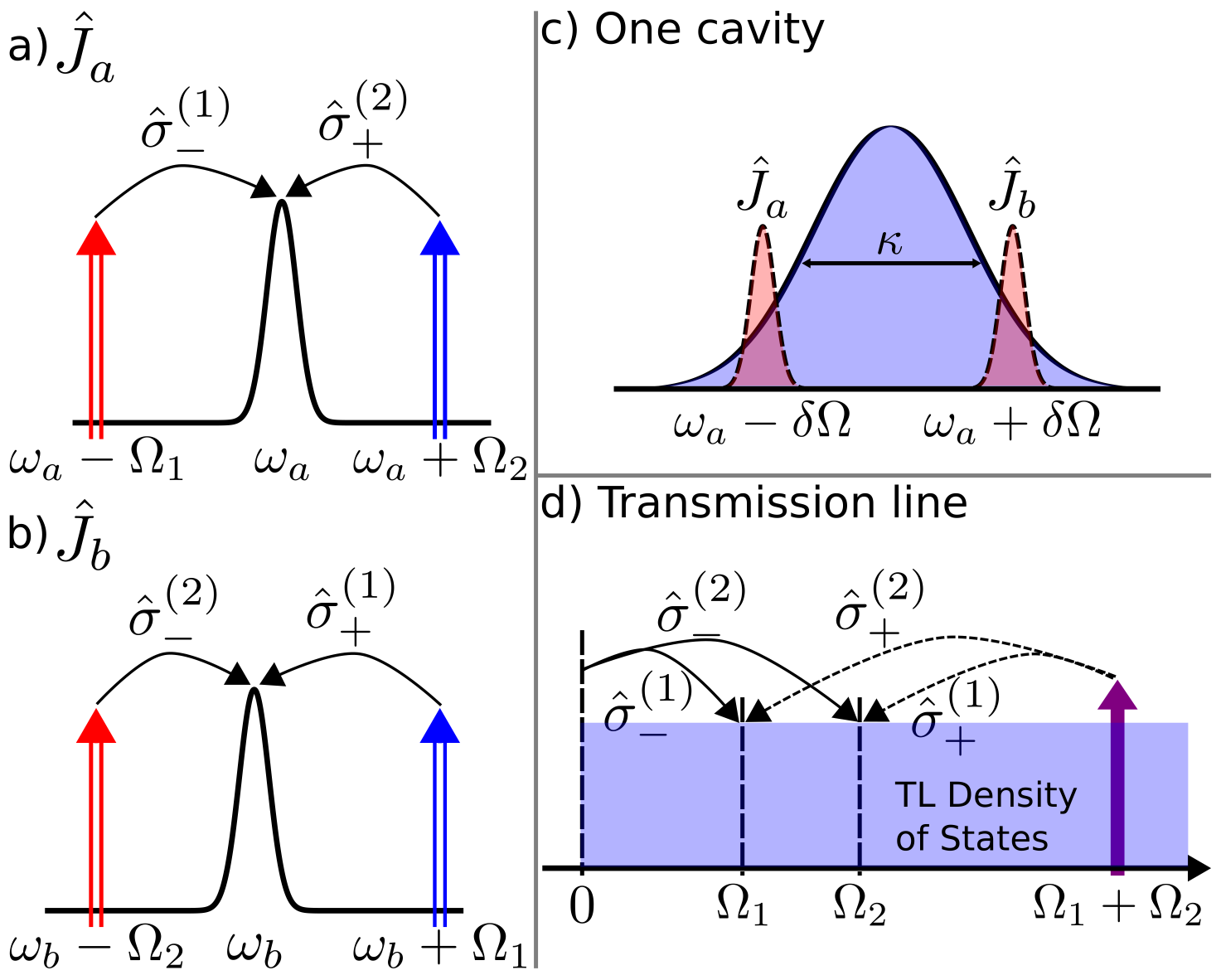}
  \caption{{\bf Two-qubit interference.} {\bf a)} and {\bf b)} interference between the qubit decay and excitation processes mediated by the coupling modulation that results in the combined dissipators $\hat{J}_{a/b}$. {\bf c)} For a cavity with a broad linewidth, the two dissipative qubit channels sample the cavity density of states at different frequencies, and effectively interact with distinct environments. {\bf d)} In the transmission line setup, due to the modulated coupling both qubit 1 (2) decay and qubit 2 (1) excitation result in a TL excitation at $\Omega_1$ ($\Omega_2$).}
  \label{fig:2Q}
\end{figure}
%%%%%%%%%

The above ideas can easily be extended to two qubits (frequencies $\Omega_1$ and $\Omega_2$), allowing one to mimic the effects of a two-mode squeezed reservoir; we refer to this general idea as synthetic two-mode squeezing. We now use phase-locked coupling modulation between the qubits and two independent decay channels (mediated by bosonic modes). As in the single qubit case, the key ingredient is interference between red/blue sideband Raman processes that result in photon emission at the same frequency. However, in the two-qubit setup the red sideband for one qubit and the blue sideband for the other interfere, as depicted in Fig.~\ref{fig:2Q}a) and b). This results in the reduced qubit-only master equation
\begin{align}
&\dot\hrho_q = \frac{4\bar{g}^2}{\kappa}\Big(\mathcal{D}\left[\hat{\mathcal{J}}_a\right] + \mathcal{D}\left[\hat{\mathcal{J}}_b\right]\Big)\hrho_q, \label{eqn:ME2}
\end{align}
with
\begin{align}
  \hat{\mathcal{J}}_a = \al_{-}\hat{\sigma}^{(1)}_- + \al_{+}\hat{\sigma}^{(2)}_+,~~\hat{\mathcal{J}}_b = \beta_{+}\hat{\sigma}^{(1)}_+ + \beta_{-}\hat{\sigma}^{(2)}_-,
\end{align}
describing the independent qubit dissipative channels mediated by the bosonic modes shown in Fig.~\ref{fig:2Q}a) and 3b). Here $\bar{g}$ and $\alpha_{\pm}$ ($\beta_{\pm}$) define the qubit-mode coupling modulation for mode $a$ ($b$), in direct analogy to Eq.~\eqref{eqn:coup}. By setting $\al_{-} = \beta_{-} = \alpha\cosh(r)$ and $\al_{+} = \beta_{+} = \alpha\sinh(r)$ we obtain Eq.~\eqref{eqn:qTMS} with $\gamma_1=\gamma_2 = 4\bar{g}^2\alpha^2/\kk$. We now present three setups to engineer Eq.~\eqref{eqn:ME2}, leaving detail of the calculations to App.~\ref{app:2Q}, .

The first and simplest setup consists of two cavities, each coupled to both qubits, as described by the coupling Hamiltonian
\begin{align}
\hat{H}_{\rm coup} = \left[g_1(t)\left(\hat{a}+\hat{a}^\dagger\right) + g_2(t)\left(\hat{b}+\hat{b}^\dagger\right)\right]\left(\hat{\sigma}^{(1)}_x + \hat{\sigma}^{(2)}_x\right).
\end{align}
The coupling $g_1(t)$ is modulated such that the $a$ cavity mediates the interference process of Fig.~\ref{fig:2Q}a) described by the dissipator $\hat{\mathcal{J}}_a$, and similarly $g_2(t)$ and the $b$ cavity mediate Fig.~\ref{fig:2Q}b) and $\hat{\mathcal{J}}_b$. The $a$ and $b$ modes need not be spatially localized:  they could, e.g.~,~be the hybridized supermodes of two tunnel-coupled physical cavities, with the underlying localized cavity modes coupling to only one qubit each. If instead we interfere both red-sideband processes and both blue-sideband processes, as can be done with trivial modification of this setup, we would obtain the dissipators
\begin{align}
  \hat{\mathcal{J}}^{f}_a = \al_{-}\hat{\sigma}^{(1)}_- + \al_{+}\hat{\sigma}^{(2)}_-,~~ \hat{\mathcal{J}}^{f}_b = \beta_{+}\hat{\sigma}^{(1)}_+ + \beta_{-}\hat{\sigma}^{(2)}_+, \label{eqn:qTMSflip}
\end{align}
which leads to Eq.~\eqref{eqn:qTMS} with the state of qubit 2 flipped.

The second setup uses a single cavity with a broad linewidth ($\kappa$) to supply the two distinct dissipative channels for the qubits, as depicted in Fig.~\ref{fig:2Q}c), with the cavity coupling to both qubits with the same modulated coupling, given by the Hamiltonian
\begin{align}
\hat{H}_{\rm coup} = g(t)\left(\hat{a}+\hat{a}^\dagger\right)\left(\hat{\sigma}^{(1)}_x + \hat{\sigma}^{(2)}_x\right).
\end{align}
In this setup, as a result of the modulated coupling, one pair of sideband processes results in cavity photon emission at frequency $\omega_a + \delta\Omega$, and the other pair results in emission at $\omega_a - \delta\Omega$, with $\delta\Omega = (\Omega_1-\Omega_2)/2$. These two mediated decay processes are independent and add incoherently, which leads to Eq.~\eqref{eqn:ME2}. Note that this setup cannot easily be modified to directly engineer Eq.~\eqref{eqn:qTMSflip}.

Finally, the third setup is two qubits connected to a long waveguide or transmission line (TL) with static and modulated coupling, as depicted in Figs.~\ref{fig:setup}b) and \ref{fig:2Q}d). Modulation of qubit 1 coupling at the qubit sum-frequency drives a Raman process where qubit 1 is excited and a photon of frequency $\Omega_{2}$ is emitted into the TL. Due to the static coupling, qubit 2 also decays into the TL at frequency $\Omega_{2}$. These two processes interfere, as do the analogous processes at emission frequency $\Omega_1$ when the qubits' roles are reversed, which results in qubit evolution described by Eq.~\eqref{eqn:ME2}. While this setup is sensitive to the distance between the qubits (as we do not want a waveguide-mediated Hamiltonian qubit-qubit interaction, see App.~\ref{app:2QTL}), it can be used to generate and stabilize steady-state entanglement between two remote qubits.
We note that several recent circuit QED experiments studied setups with tuneable qubit-transmission line couplings \cite{Zhong:aa,ClelandPRL2020}, a platform that could be ideal for the above implementation.

%%%%%%%%%%%%%%%%%%%%%%%%%%%%%
%%%%%%%%%%%%%%%%%%%%%%%%%%%%%

\section{Two-Qubit Synthetic Squeezed Dissipation via local linear driving and collective loss}
\label{sec:collective-loss}

The idea of generating remote entanglement by synthetic two-mode squeezing is appealing; however, not all architectures facilitate dynamically modulated qubit-waveguide couplings.  In this section, we show that one can still use synthetic squeezing ideas to stabilize entanglement by simply combining the passive, collective loss generated by a photonic link with tailored local qubit Rabi drives.  In the correct reference frame, the physics of this setup is almost directly analogous to driving two qubits with two-mode squeezed light (c.f.~Eq.~(\ref{eqn:qTMS})).  Our discussion here both generalizes and helps provide intuition for the dissipative entanglement protocol presented in Ref.~\cite{Motzoi:2016aa}.

Consider two qubits which are coupled to a common source of loss at a rate $\Gamma$.
The dissipation is described by the master equation 
$\dot\hrho_q = \Gamma \mathcal{D}\left[\hat{\mathcal{J}}\right] \hrho_q $
with the collective loss operator
\begin{equation}
    \hat{\mathcal{J}} = \hat{\sigma}_{-}^{(1)} + \eta\hat{\sigma}_{-}^{(2)},
    \label{eq:collective-jump}
\end{equation}
where $\eta$ allows for an asymmetry between the qubit-bath couplings.  For clarity we take $\eta=1$ in what follows; as we show later, analogous results hold for $\eta \neq 1$.  $\hat{\mathcal{J}}$ has a two-parameter family of dark states of the form 
\begin{equation}
    \ket{\Phi[\alpha,\phi]} = \sqrt{1-\alpha^2}\ket{00}+\frac{e^{i\phi}\alpha}{\sqrt{2}}(\ket{01}-\ket{10}),
    \label{eq:dark-family}
\end{equation}
where $0\leq\alpha\leq 1$.  These states are entangled for any $\alpha>0$, whereas for $\alpha=0$ we have the trivial dark state $\ket{00}$.  Any state (pure or impure) in the span of this dark state manifold is a possible steady state, making this dissipative evolution of little use if the goal is entanglement stabilization.  As such, we would like to introduce additional simple dynamics to the qubits so that there is a {\it unique} entangled dark state (for some $\alpha >0$).  
% The question is how best to do this.  

We will solve this problem by making a direct mapping to entanglement via two-mode squeezed dissipation.  This is possible because the dark states in Eq.~(\ref{eq:dark-family}) are unitarily equivalent to the ``paired'' family of entangled states associated with two-mode squeezed dissipation, i.e.~states of the form in Eq.~\eqref{eqn:EntQ}.  Without loss of generality we set $\phi=0$; in this case
we have the explicit mapping
\begin{align}
    \ket{\Psi[\alpha]}_q &\equiv 
        \hat{U}[\alpha] \ket{\Phi[\alpha]} \nonumber \\
        & = 
    \frac{1}{\sqrt{\cosh (2r)}}\left( \cosh(r)\ket{00} - \sinh(r)\ket{11} \right),
    \label{eq:entangled-family}
\end{align}
where $r[\alpha]$ is determined via $\alpha = \sqrt{\tanh(2r)}$.  Crucially, the required unitary is {\it local}, i.e. $\hat{U}[\alpha] = \hat{U}^{(1)}[\alpha]\otimes\hat{U}^{(2)}[\alpha]$.  It corresponds to opposite local rotations of each qubit about the y-axis
\begin{align}
    \hat{U}^{(j)}[\alpha] = \exp(\pm i\frac{\theta}{2} \hat{\sigma}_y^{(j)}),
    \label{eq:local-unitary}
\end{align}
where qubit 1 (2) takes the upper (lower) sign. The rotation angle $\theta$ is a function of $\alpha$ (or equivalently $r$), and is given by
\begin{equation}
    \cos\theta = \sqrt{\frac{1-\alpha^2}{1+\alpha^2}} = e^{-2r}.
    \label{eq:transform-angle}
\end{equation}
Picking out a non-trivial dark state is now mapped in the new frame to ensuring the steady state has a particular non-zero value of the squeezing parameter $r$.  This latter task is something we know how to do, namely by using the synthetic squeezing master equation in Eq.~(\ref{eqn:qTMS}).

In what follows, we choose a particular value of $r=r_0$, corresponding to the entangled state we would like to stabilize; we will also view $\hat{U}$ and $\ket{\Psi}_q$ to be functions of $r$ rather than $\alpha$.  The next step in our mapping to synthetic two-mode squeezing is to examine the form of the collective loss dissipator $\hat{\mathcal{J}}$ in the new frame defined by $\hat{U}[r_0]$.
We have
\begin{equation}
    \hat{\mathcal{J}}' \equiv 
    \hat{U}[r_0] \hat{\mathcal{J}} \hat{U}^{\dagger}[r_0]
    =  \hat{\mathcal{J}}'_{1} + \hat{\mathcal{J}}'_{2} + \hat{\mathcal{J}}'_{Z}.
    \label{eq:xfrm-coll-loss}
\end{equation}
Here, $\hat{\mathcal{J}}_{1}'$ and $\hat{\mathcal{J}}_{2}'$ are precisely the two squeezing dissipators needed for synthetic squeezing (c.f.~Eq.~\ref{eqn:qTMS}))
\begin{align}
    \hat{\mathcal{J}}_{1}^{\prime}&=e^{-r_0}\left(\cosh(r_0)\hat{\sigma}_{-}^{(1)}-\sinh(r_0)\hat{\sigma}_{+}^{(2)}\right),\label{eq:coll-loss-sq1}\\
    \hat{\mathcal{J}}_{2}^{\prime}&=e^{-r_0}\left(\cosh(r_0)\hat{\sigma}_{-}^{(2)}-\sinh(r_0)\hat{\sigma}_{+}^{(1)}\right).\label{eq:coll-loss-sq2}
\end{align}
Further, $\hat{\mathcal{J}}_{Z}'$ is a correlated dephasing term
\begin{equation}
    \hat{\mathcal{J}}_{Z}' = \frac{1}{2}\sqrt{1-e^{-4r_0}} \left( \hat{\sigma}_{z}^{(1)} - \hat{\sigma}_{z}^{(2)} \right).
    \label{eq:coll-loss-dp}
\end{equation}

Hence, in the new frame defined by our chosen value of $r=r_0$, the original collective loss dissipator has a direct connection to the synthetic squeezing dissipators in Eq.~(\ref{eqn:qTMS}):  $\mathcal{J}'$ is the {\it sum} of these dissipators, plus a dephasing dissipator.  One can easily check that the paired target state $\ket{\Psi[r_0]}_q$ is dark with respect to this sum dissipator.  However, it is not unique, which is to be expected:   all  we have done at this stage is re-written our original problem in a new frame, hence the transformed version of any dark state from Eq.~(\ref{eq:dark-family}) (including $\ket{00}$) remains a dark state.

However, in our new frame defined by $\hat{U}[r_0]$, the state $\ket{\Psi[r_0]}_q$ is the {\it only} dark state of the dissipator $\hat{\mathcal{J}}'$ that does not have any single-excitation components, i.e.~kets $\ket{01}$ or $\ket{10}$.  This suggests a very simple strategy to make the paired state $\ket{\Psi[r_0]}_q$ a unique dark state. By adding a Hamiltonian that (in the new frame) breaks the degeneracy between the paired and unpaired subspaces, such as 
\begin{equation}
    \hat{H}' = \frac{\mu}{2}\left( \hat{\sigma}_{z}^{(1)} - \hat{\sigma}_{z}^{(2)} \right),
    \label{eq:xfrm-H-coll-loss}
\end{equation}
we introduce an energy gap $\mu$ between the paired states $\ket{00}$ and $\ket{11}$ (which have zero energy) and the unpaired states $\ket{10}$ and $\ket{01}$. 

The above extra Hamiltonian dynamics is exactly what we need to pick out a unique entangled steady state.  One can directly confirm that for any $\mu > 0$, the target entangled state $\ket{\Psi[r_0]}_q$ is the {\it unique} steady state of the dynamics generated by our master equation, which in the new frame has the form 
\begin{equation}
    \dot{\hrho}' = -i[\hat{H}',\hrho'] + 
        \Gamma \mathcal{D}\left[\hat{\mathcal{J}}'_{1} + \hat{\mathcal{J}}'_{2} + \hat{\mathcal{J}}'_{Z} \right]\hrho'.
    \label{eq:coll-loss-qme}
\end{equation}
Here, $\hrho' = \hat{U}[r_0] \hrho \hat{U}^{-1}[r_0]$ is the transformed-frame density matrix of the two qubits.  

Note that for large $\mu \gg \Gamma$, the connection to the synthetic squeezing master equation of Eq.~(\ref{eqn:qTMS}) is even more direct.  In this limit, $\hat{H}'$ will cause $\hat{\mathcal{J}}'_{1/2}$ to oscillate at frequencies $\pm\mu$, while $\hat{\mathcal{J}}'_Z$ remains static.  In this case, the dynamics is well approximated by dropping cross-terms in the dissipator (as they will be fast oscillating), leading to the approximation
\begin{equation}
    \dot{\hrho}' \simeq -i[\hat{H}',\hrho'] + 
        \Gamma \left( 
            \mathcal{D}\left[\hat{\mathcal{J}}'_{1}\right] 
        + \mathcal{D}\left[\hat{\mathcal{J}}'_{2}\right] 
        + \mathcal{D}\left[\hat{\mathcal{J}}'_{Z} \right] \right) \hrho'.
    \label{eq:coll-loss-qme2}
\end{equation}
Apart from the last collective dephasing dissipator, the dissipative dynamics is now  completely equivalent to the synthetic two-mode squeezing master equation in 
Eq.~(\ref{eqn:qTMS}). Importantly, $\ket{\Psi[r_0]}_q$ is still the unique steady state of this master equation. Moreover, as we discuss in App.~\ref{app:MS2}, this additional dephasing has little impact on the dynamics in the limit $\mu \gg \Gamma$. In this limit we can also remove the Hamiltonian term by moving to the interaction frame, $\hat{\rho}'' = \exp\small(-i\hat{H}'t\small) \hat{\rho}' \exp\small(i\hat{H}'t\small)$, where the equivalence to Eq.~\eqref{eqn:qTMS} is more obvious, and $\ket{\Psi[r_0]}_q$ still remains the unique steady state. 

Finally, having realized a master equation that is strongly analogous to synthetic two-mode squeezing in the new frame, we can move back to the original frame (i.e.~via the unitary $\hat{U}^\dagger[r_0]$).  The master equation in the original frame has the form
\begin{equation}
    \dot{\hrho} = -i[\hat{H},\hrho] + 
        \Gamma  
            \mathcal{D}\left[\hat{\mathcal{J}} \right] \hrho,
    \label{eq:coll-loss-lab}
\end{equation}
where $\hat{\mathcal{J}}$ is the simple collective loss dissipator in Eq.~(\ref{eq:collective-jump}), and the lab-frame Hamiltonian $\hat{H} = \hat{U}^{\dagger}[r_0] \hat{H}' \hat{U}[r_0]$ has the form
\begin{equation}
    \hat{H}=\frac{\Delta}{2}\left(\hat{\sigma}_{z}^{(1)}-\hat{\sigma}_{z}^{(2)}\right)+\Lambda \left(\hat{\sigma}_{x}^{(1)}+\hat{\sigma}_{x}^{(2)}\right).
    \label{eq:coll-diss-H-og-frame}
\end{equation}
with 
\begin{align}
    \Delta  = \mu e^{-2r_0},  \,\,\,\,\, 
    \Lambda = (\mu/2) \sqrt{1-e^{-4r_0}}.  
    \label{eq:HDriveParams}
\end{align}

This Hamiltonian describes a two-qubit system subject to Rabi drives at the same frequency $\nu$ and amplitude $\Lambda$, in the rotating frame determined by the drive frequency. Thus, $\pm \Delta$ describe the detuning of the drive from each qubit, such that $\nu$ is the average of the two qubit frequencies.  To be explicit, in the lab frame where the drives are time-dependent, the Hamiltonian has the form
\begin{equation}
    \hat{H}_{\rm lab} = \frac{1}{2} \left( \Omega_{1} \hat{\sigma}_{z}^{(1)}  + 
    \Omega_{2} \hat{\sigma}_{z}^{(2)} \right) +
    \Lambda \left[ e^{i \nu t} \left( \hat{\sigma}_{-}^{(1)} +
    \hat{\sigma}_{-}^{(2)} \right) + \textrm{h.c.} \right]
\end{equation}
with $\nu= (\Omega_1 + \Omega_2) / 2$ and $\Delta = \Omega_1 - \Omega_2$.

We note that the master equation in Eq.~(\ref{eq:coll-loss-lab}) was first presented in Ref.~\cite{Motzoi:2016aa}.  Our derivation above adds to this work by showing the connection to squeezed dissipation, and by providing a clear intuitive explanation for why the scheme stabilizes entanglement.  Further, our approach also makes it easy to generalize to the case where the qubits do not couple equally to the collective loss channel 
(i.e.~$\eta\neq 1$ in Eq.~\eqref{eq:collective-jump}), something not previously considered.  The asymmetric case is presented in detail in App.~\ref{app:MS}, and the key finding is that dissipative entanglement stabilization is still possible, but the parameters in the required driving Hamiltonian are different:
\begin{equation}
    \hat{H} = \frac{\Delta+\epsilon}{2}\hat{\sigma}_{z}^{(1)}-\frac{\Delta-\epsilon}{2}\hat{\sigma}_{z}^{(2)}+\Lambda\left(\eta\hat{\sigma}_{x}^{(1)}+\hat{\sigma}_{x}^{(2)}\right),
    \label{eq:asymmetricloss-H}
\end{equation}
where
\begin{equation}
    \epsilon = \frac{\Lambda^{2}}{\Delta}\left(1-\eta^{2}\right),
\end{equation}
and where $\Delta,~\Lambda$ are no longer given by Eq.~(\ref{eq:HDriveParams}) but are determined via Eqs.~\eqref{eq:rabi-detuning-general} and \eqref{eq:rabi-drive-general} in App.~\ref{app:MS}.

While the above scheme is attractive as it does not require the ability to modulate the qubit-waveguide coupling, it does have other limitations when compared to the modulated-coupling approach.  For the symmetric $\eta=1$ case and a large squeezing parameter $r$, the Rabi-drive amplitude scales as $\Lambda \sim e^{2r} (\Omega_1 - \Omega_1)/2$.  Hence, for qubits that are far detuned from one another, one requires extremely large Rabi drive amplitudes.  In contrast, the modulated-coupled approach of Sec.~\ref{sec:2qubit} has no analogous limitation. 

Finally, we briefly note some of the imperfections of the above scheme. The collective loss dissipator is typically realized by coupling the qubits to a waveguide.  When the separation of the qubits along the waveguide is correctly set, only collective loss dissipation is generated as desired. However, if the qubit spacing is not equal to one of the special values, the waveguide will also generate single qubit loss dissipators, as well as a coherent tunneling term $\hat{H}_{\rm wg} = J(\hat{\sigma}_+^{(1)}\hat{\sigma}_-^{(2)} + {\rm h.c.})$. As discussed in Ref.~\cite{Motzoi:2016aa}, single qubit loss rapidly degrades the entanglement of the steady state. The coherent tunneling term will also degrade the steady state entanglement, however, one can show numerically the steady state entanglement is much more sensitive to single qubit loss than to the tunneling term. Intuitively, this is because one of the eigenstates of $\hat{H}_{\rm wg}$ is the $\alpha= 1$ dark state of the family of Eq.~\eqref{eq:dark-family} (equivalently the $r\to\infty$ two-qubit squeezed state). The modulated-coupling scheme involving a waveguide discussed in detail in App.~\ref{app:2QTL} shows an analogous result.

%%%%%%%%%%%%%%%%%%%%%%%%%%%%%
%%%%%%%%%%%%%%%%%%%%%%%%%%%%%

\section{Two-qubit Entanglement Stabilization}
\label{sec:SSE}

As we have discussed, the dynamics of Eq.~\eqref{eqn:qTMS} can be engineered without squeezed vacuum, either using our proposed coupling modulation schemes (see App.~\ref{app:2Q} for further details), or the linear driving and correlated loss scheme of Ref.~\cite{Motzoi:2016aa} that we have generalized in the previous section (details in App.~\ref{app:MS}). Thus, it is useful to understand the limits of using Eq.~\eqref{eqn:qTMS} for high-quality entangled state stabilization, as it is the generic description of many schemes.

The steady-state of this evolution is the entangled state of Eq.~\eqref{eqn:EntQ}, which for $r\rightarrow \infty$ tends to a maximally entangled Bell state. Our implementation of Eq.~\eqref{eqn:ME2} does not require increasing modulation amplitudes to increase $r$, which would otherwise be an issue given the RWA used to drop fast-oscillating time-dependent terms from Eq.~\eqref{eqn:ME2}. For increasing $r$, $\alpha$ can be decreased to keep $\al_{\pm}$ and $\beta{\pm}$ constant. As in the single qubit case, this comes at the cost of reducing the overall decay rate $\Gamma \equiv 4\bar{g}^2\alpha^2/\kk$. Keeping the modulation amplitudes fixed, more entanglement requires a longer preparation time, a trade-off also observed in similar bosonic schemes \cite{Woolley:2014aa}.

Moreover, saturating the $r\rightarrow \infty$ limit leads to a degeneracy in the steady-state of Eq.~\eqref{eqn:qTMS} (see App.~\ref{app:Balance} for further details). This degeneracy can be most easily understood by considering dynamics in the $r\rightarrow \infty$ limit of  Eq.~\eqref{eqn:qTMSflip}
\begin{align}
  \dot\hrho_q &= \gamma\left(\mathcal{D}\left[\hat{\sigma}_1^- + \hat{\sigma}_2^-\right] + \mathcal{D}\left[\hat{\sigma}_1^+ + \hat{\sigma}_2^+\right]\right)\hrho_q, \label{eqn:qTMSv2}
\end{align}
which is unitarily equivalent to Eq.~\eqref{eqn:qTMS}. In this form, it is clear that the steady-state degeneracy is due to the evolution conserving total angular momentum of the two qubits (implying the existence of distinct singlet and triplet steady states).  
In light of this, for $r$ large but not infinite, we expect a slow approach to steady-state due to the weak breaking of total angular momentum conservation. This slow dynamical rate is intrinsic to Eq.~\eqref{eqn:qTMS}, as opposed to the slow decay rate discussed in the previous paragraph, which is an effect of our engineering of Eq.~\eqref{eqn:ME2}.

\subsection{Slow Dissipation: Liouvillian Gap and Large-Spin Mean-Field}

We now focus on the intrinsically slow dynamics of Eq.~\eqref{eqn:qTMS}, which correspond to decay out of the nearly degenerate subspace discussed in App.~\ref{app:Balance}. Writing $\dot{\hrho}_q = \mathcal{L}(\hrho_q)$, where $\mathcal{L}$ corresponds to Eq.~\eqref{eqn:ME2}, we can vectorize this equation as $\dot{\vec{\rho}}_q = \hat{\mathcal{L}}\vec{\rho}_q$, where the super-operator matrix $\hat{\mathcal{L}}$ describes the dissipative evolution of the qubits. The eigenvalues of this matrix describe the dynamical timescales of the system evolution, and we find that there is one slow rate, corresponding to the eigenvalue with the smallest real part, often referred to as the Liouvillian gap. In the large $r$ limit it scales inversely with $r$ as
\begin{align}
  \kappa_{\rm slow} = \frac{\Gamma}{3\sinh^2(r)} + \mathcal{O}\left(\frac{1}{\sinh^4(r)}\right).
  \label{eq:kappa-slow}
\end{align}
Note that all other eigenvalues monotonically increase with $r$, scaling as $\sinh^2(r)$ to leading order.

This intrinsically slow rate implies that there is a trade-off between the stabilization speed-limit and the amount of entanglement in steady-state (cf.~Eq.~\eqref{eqn:EntQ}). The impact on a specific protocol depends on whether ``typical'' initial conditions evolve to the steady-state via the process described by this slow rate. We leave this question to future work, but note that prior work has shown that Eq.~\eqref{eqn:qTMS} is quite robust to additional sources of qubit decoherence \cite{Kraus:2004aa}, which ultimately compete with this slow rate and determine the asymptotic entanglement.

This slow rate is a property of the two qubit setup, and does not exist for its bosonic analogue \cite{Woolley:2014aa}, where $\hat{\sigma}_{1/2}^-$ are replaced by bosonic modes $\hat{d}_{1/2}$, as the bosonic model has a unique steady-state for balanced dissipators. To understand the transition between these two extremes, we consider a system of two spin-$S$ particles, where $S\gg~1/2$. The bosonic model is often used as an approximation to such a large spin model via the Holstein-Primakoff (HP) transformation \cite{HolsteinPrimakoff}, which connects the spin operator $\hat{S}^+$ to a bosonic mode via $\hat{S}^+ = \hat{d}^\dagger\sqrt{2S - \hat{d}^\dagger\hat{d}}$. In the HP representation, the large spin version of Eq.~\eqref{eqn:ME2} has dissipators
\begin{align}
    &\hat{\mathcal{J}}_a = \cosh(r)\sqrt{1 - \frac{\hat{n}_1}{2S}}~\hat{d}_1 + \sinh(r)\hat{d}_2^\dagger\sqrt{1 - \frac{\hat{n}_2}{2S}}, \label{eqn:LS1} \\
    &\hat{\mathcal{J}}_b = \cosh(r)\sqrt{1 - \frac{\hat{n}_2}{2S}}~\hat{d}_2 + \sinh(r)\hat{d}_1^\dagger\sqrt{1 - \frac{\hat{n}_1}{2S}}, \label{eqn:LS2}
\end{align}
with $\hat{n}_k = \hat{d}_k^\dagger\hat{d}_k$, and an enhanced dissipative rate $\Gamma_S \equiv 16S\bar{g}^2\alpha^2/\kk$. In the limit $S\gg \left<\hat{n}_{k}\right>$ the square root factors can be neglected and the bosonic model is obtained.

However, the steady-state of the bosonic model has $\left<\hat{n}_{k}\right> = \sinh^2(r)$, and the square root factor of Eqs.~\eqref{eqn:LS1} and \eqref{eqn:LS2} cannot be ignored. Treating this factor in the mean-field limit, we obtain a scalar that can be absorbed into an effective asymptotic dissipative rate for the large spin model
\begin{align}
    \Gamma^{\rm MF}_S = \frac{16S\bar{g}^2\alpha^2}{\kk}\left(1-\frac{\sinh^2(r)}{2S}\right).
\end{align}
This rate slows as $r$ increases, similar to what was observed in the two-qubit ($S=1/2$) model. Dynamically, $\Gamma^{\rm MF}_S$ will decrease as $\left<\hat{n}_{k}\right>$ increases during the protocol. This mean-field result is a strong indication that any finite spin model may have a slow rate that results in a trade-off between the speed and quality of entanglement stabilization. Further study, beyond mean-field, is warranted to confirm this result, but is outside the scope of this work.

\subsection{Thermal Dissipative Squeezing}
\label{sec:therm}
%%%%%%%%
%Fig. 4
%%%%%%%%
\begin{figure}[t]
  \includegraphics[width=0.65\columnwidth]{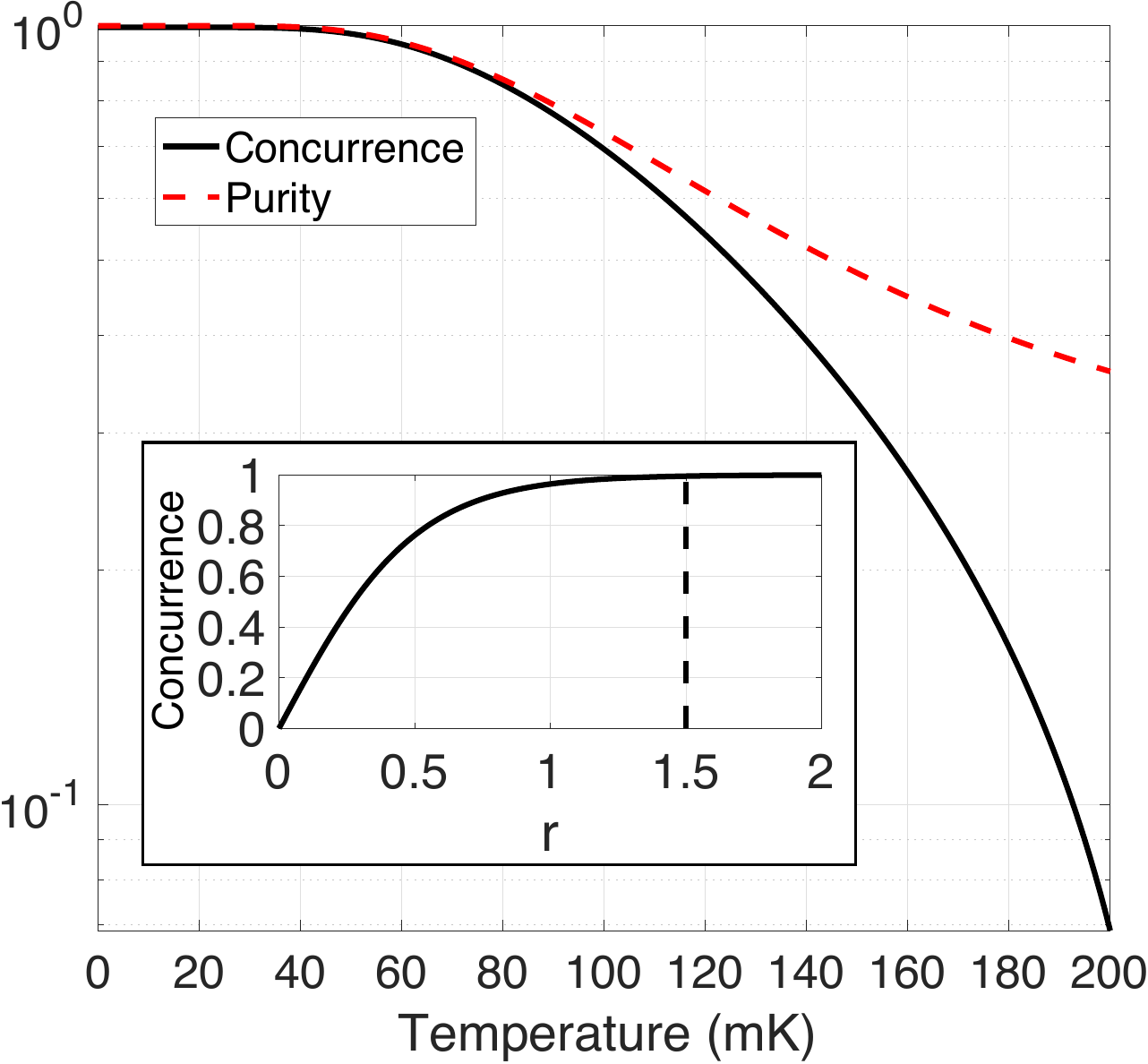}
  \caption{{\bf Thermal dissipative squeezing.} Concurrence (black solid line) and purity (red dashed line) of the steady-state of thermal dissipative squeezing, cf.~Eq.~\eqref{eqn:TDS}, as a function of the temperature of the transmission line. The inset shows the concurrence as a function of the squeezing parameter $r$ for zero temperature, and the dashed horizontal line indicates the squeezed parameter value used in the main plot.}
  \label{fig:Con}
\end{figure}
%%%%%%%%%

The effect of thermal occupation of the bosonic environment of Eq.~\eqref{eqn:ME2} is especially relevant in the transmission line version of our scheme, and remains unexplored in prior literature. To study this, we consider the thermal dissipative squeezing master equation
\begin{align}
  \nonumber\dot\hrho_q =~&\Gamma\Big(\left(1+\bar{n}_{\rm th}\right)\mathcal{D}\left[\hat{\mathcal{J}}_a\right] + \bar{n}_{\rm th}\mathcal{D}\left[\hat{\mathcal{J}}^\dagger_a\right] \\ & + \left(1+\bar{n}_{\rm th}\right)\mathcal{D}\left[\hat{\mathcal{J}}_b\right] + \bar{n}_{\rm th}\mathcal{D}\left[\hat{\mathcal{J}}^\dagger_b\right]\Big)\hrho_q, \label{eqn:TDS}
\end{align}
with $\hat{\mathcal{J}}_a$ and $\hat{\mathcal{J}}_b$ as before for $\al_{-} = \beta_{-} = \alpha\cosh(r)$, $\al_{+} = \beta_{+} = \alpha\sinh(r)$, and where $\bar{n}_{\rm th}$ is the average photon population of the modes.

We characterize the effect of thermal occupation by measuring the concurrence of the two-qubit steady-state as a function of transmission line temperature, with the thermal occupation as a function of temperature given by the Bose-Einstein distribution. For qubits with a frequency around 6 GHz (as is typical in circuit QED) this is shown in Fig.~\ref{fig:Con}. As can be seen, the two-qubit steady-state remains highly entangled ($>0.9$ concurrence) up to a temperature of around $70$ mK. For comparison, we also plot the purity of the steady-state as a function of temperature. While 70 mK is well above the operating temperature of many cryogenic circuit QED experiments, the relevant quantity is the thermal occupation of the TL near the qubit frequency, which includes thermal noise from classical control lines. If this control line noise is controlled via appropriate filtering \cite{Wang2019}, high-quality remote entanglement can be generated by our TL scheme.

\section{Conclusion}
\label{sec:conc}

In this paper, we have presented an overlooked connection linking the dynamics of a system of qubits immersed in a squeezed environment, and reservoir engineering using interference of qubit decay and excitation pathways. We have shown that both situations result in the exact same dissipative dynamics, and have proposed schemes which achieve this interference using parametric coupling modulation of the Jaynes-Cummings interaction of cavity- and waveguide-QED. The latter can be used to remotely entangle qubits coupled to a transmission line.

Such modulated coupling is readily available in circuit-QED \cite{Chen:2014aa,McKay:2016aa,Lu:2017aa,Zhong:aa,ClelandPRL2020}, and our scheme could immediately be implemented in such architectures. An alternative implementation of the modulated coupling of our schemes uses a dispersive interaction with classical cavity displacement \cite{Hacohen-Gourgy:2016aa,Eddins:2018aa}. In either case, the bare-coupling can be very weak, as the amplitude of the modulation can be used to compensate. Adapting to the specifics of a particular setup, and understanding the impact of spurious decoherence sources will be the focus of future work.

The results of our work overturn the long held belief that the unique effects of qubit-squeezed vacuum interaction were a feature of the nonclassical nature of the squeezed environment. In effect, we have traded the nonclassical radiation for a modulated nonlinear coupling. This combination of nonlinearity and classical modulation can be compared to the operation of a parametric amplifier, whose nonlinearity is provided by one or more two-level systems (qubits) coupled to a bosonic mode, such as in a Josephson parametric amplifier \cite{Yurke:1989aa,Castellanos-Beltran:2008aa}. In the parametric amplifier it is the bosonic field that is squeezed, while in our scheme, the bosonic mode is never squeezed, but rather the qubit undergoes the dissipative dynamics corresponding to a squeezed environment.

We note that schemes for dissipative Bell state stabilization involving two qubits, one or more coupled cavities, and classical qubit-cavity control have been proposed and implemented in circuit-QED \cite{Leghtas:2013aa,Shankar:2013aa,Aron:2014aa,Kimchi-Schwartz:2016aa,Motzoi:2015aa,Motzoi:2016aa,Doucet:aa,Brown2021}. Our work is the only scheme to make explicit connection to dissipative stabilization using squeezed vacuum, and it remains to be seen if this connection can be made for other proposals that use classical drives or coupling modulation. We conjecture that this connection holds approximately for all schemes with an intrinsically slow rate, as we have shown for the scheme of Ref.~\cite{Motzoi:2016aa}. For schemes with no slow rate, such as Ref.~\cite{Doucet:aa}, understanding where the connection to dissipative squeezing breaks is a topic worthy of future study.

\acknowledgements
This material is based upon work for which LG was supported by the U.S. Department of Energy, Office of Science, under Award Number DE-SC0019461, and for which AL and AC were supported by the National Science Foundation QLCI for HQAN (NSF award 2016136).

\bibliography{SynSqzBib}

\appendix

\clearpage

\section{Two Qubit Master Equation Derivations}
\label{app:2Q}

\subsection{Two Qubits and Two Cavities}
\label{app:2Q2C}

An instructive example is to directly extend the single qubit setup to two qubits, by considering two qubits and cavities coupled via the Hamiltonian
\begin{align}
\nonumber\hat{H} &= \omega_a\hat{a}^\dagger\hat{a} + \frac{\Omega_1}{2}\hat{\sigma}^{(1)}_z  +g_1(t)\left(\hat{a}+\hat{a}^\dagger\right)\left(\hat{\sigma}^{(1)}_x + \hat{\sigma}^{(2)}_x\right) \\ &+ \omega_b\hat{b}^\dagger\hat{b} + \frac{\Omega_2}{2}\hat{\sigma}^{(2)}_z + g_2(t)\left(\hat{b}+\hat{b}^\dagger\right)\left(\hat{\sigma}^{(1)}_x + \hat{\sigma}^{(2)}_x\right),  \label{eq:Hint2Q}
\end{align}
with the modulated couplings
\begin{align}
  &\nonumber g_1(t) = \bar{g}\left(\al_{-}e^{-i\left(\omega_a - \Omega_1\right) t} + \al_{+}e^{-i\left(\omega_a + \Omega_2\right) t}\right) + h.c.,\\
  &\nonumber g_2(t) = \bar{g}\left(\beta_{-}e^{-i\left(\omega_b - \Omega_2\right) t} + \beta_{+}e^{-i\left(\omega_b + \Omega_1\right) t}\right) + h.c..
\end{align}
Both qubits must couple to both cavities, to allow for the interference required by dissipative squeezing. However, the symmetry in the qubit-cavity couplings shown above is not required, and can be compensated for by the modulation amplitudes. Besides the obvious physical implementation of two qubits both coupled to each physical cavity, this setup could also be implemented in a scheme where each qubit couples to only one physical cavity, and the cavities are tunnel coupled. The cavity supermodes would then play the role of the modes $\hat{a}$ and $\hat{b}$ in Eq.~\eqref{eq:Hint2Q}.

Following the same procedure as for one qubit, we move to the interaction frame for Eq.~\eqref{eq:Hint2Q} and drop all fast-oscillating terms to obtain
\begin{align}
\nonumber \hat{H}' &= \bar{g}\hat{a}^\dagger\left(\al_+\hat{\sigma}^{(2)}_{+} + \al_-\hat{\sigma}^{(1)}_{-}\right) + h.c.  \\ &+ \bar{g}\hat{b}^\dagger\left(\beta_+\hat{\sigma}^{(1)}_{+} + \beta_-\hat{\sigma}^{(2)}_{-}\right) +h.c.. \label{eqn:Hp2Q}
\end{align}
where we have assumed that $\abs{\Omega_1-\Omega_2} \gg \bar{g}$ so that all time-dependent terms are fast oscillating and can be neglected. We can eliminate the cavities to obtain an effective master equation for the qubits when $\kk_a,\kk_b \gg \abs{\al_\pm} \bar{g}, \abs{\beta_\pm} \bar{g}$. In the case where $\kk_a = \kk_b = \kk$, this gives
\begin{align}
&\dot\hrho_q = \frac{4\bar{g}^2}{\kappa}\Big(\mathcal{D}\left[\hat{\mathcal{J}}_a\right] + \mathcal{D}\left[\mathcal{J}_b\right]\Big)\hrho_q,
\end{align}
with
\begin{align}
  &\hat{\mathcal{J}}_a = \al_{-}\hat{\sigma}^{(1)}_- + \al_{+}\hat{\sigma}^{(2)}_+, \\
  &\hat{\mathcal{J}}_b = \beta_{+}\hat{\sigma}^{(1)}_+ + \beta_{-}\hat{\sigma}^{(2)}_-,
\end{align}
as depicted in Fig.~\ref{fig:2Q}a) and b). Starting with equal coupling $\bar{g}$, and setting $\kk_a = \kk_b = \kk$ is a completely general choice, as differences in the couplings or decay rates can be compensated for by control of the modulation amplitudes.

We note that if instead we modulate with red-sideband tones only on mode $\hat{a}$, and blue sideband tones on mode $\hat{b}$, i.e.
\begin{align}
  &\nonumber g_1(t) = \bar{g}\left(\al_{-}e^{-i\left(\omega_a - \Omega_1\right) t} + \al_{+}e^{-i\left(\omega_a - \Omega_2\right) t}\right) + h.c.,\\
  &\nonumber g_2(t) = \bar{g}\left(\beta_{-}e^{-i\left(\omega_b + \Omega_2\right) t} + \beta_{+}e^{-i\left(\omega_b + \Omega_1\right) t}\right) + h.c..
\end{align}
then by a similar procedure we would obtain the dissipators of Eq.~\eqref{eqn:qTMSflip}. Unfortunately, the other two approaches to engineering Eq.~\eqref{eqn:ME2} discussed in this appendix do not have a similar simple modification that would enable engineering of Eq.~\eqref{eqn:qTMSflip}.

\subsection{Two Qubits and One Cavity}
\label{app:2Q1C}

For a cavity with sufficiently broad linewidth it is also possible to synthetically generate two-qubit dissipative squeezing using our scheme with only one cavity. We start with the Hamiltonian
\begin{align}
\hat{H} &= \omega_a\hat{a}^\dagger\hat{a} + \sum_{j=1}^2\left(\frac{\Omega_j}{2}\hat{\sigma}^{(j)}_z +g(t)\left(\hat{a}+\hat{a}^\dagger\right)\hat{\sigma}^{(j)}_x\right), \label{eq:Hint2Q1C}
\end{align}
with the modulated coupling
\begin{align}
  \nonumber g(t) = \bar{g}\left(\al_{-}e^{-i\left(\omega_a - \bar{\Omega}\right) t} + \al_{+}e^{-i\left(\omega_a + \bar{\Omega}\right) t}\right) + h.c.,
\end{align}
where $\bar{\Omega} = (\Omega_1+\Omega_2)/2$ is the average qubit frequency.

We then move to the interaction frame to obtain the Hamiltonian
\begin{align}
  &\nonumber\hat{H}' = \bar{g}\hat{a}^\dagger \Big[ \\
  \nonumber &e^{i\delta\Omega t}\left(\alpha_+\hat{\sigma}_+^{(1)} + \alpha_-\hat{\sigma}_-^{(2)}\right) + e^{-i\delta\Omega t}\left(\alpha_+\hat{\sigma}_+^{(2)} + \alpha_-\hat{\sigma}_-^{(1)}\right) \\
  &\nonumber +e^{i\Sigma_1t}\alpha_-\hat{\sigma}_+^{(1)} +e^{-i\Sigma_1t}\alpha_+\hat{\sigma}_-^{(1)} \\
  &\nonumber +e^{i\Sigma_2t}\alpha_-\hat{\sigma}_+^{(2)} +e^{-i\Sigma_2t}\alpha_+\hat{\sigma}_-^{(2)} \\
  &\nonumber +e^{i(2\omega_a+\delta\Omega)t}\left(\alpha^*_+\hat{\sigma}_-^{(2)} + \alpha^*_-\hat{\sigma}_+^{(1)}\right) \\
  &\nonumber+ e^{i(2\omega_a-\delta\Omega)t}\left(\alpha^*_+\hat{\sigma}_-^{(1)} + \alpha_-^*\hat{\sigma}_+^{(2)}\right) \\
  &\nonumber+ e^{i(2\omega_a+\Sigma_1)t}\alpha^*_+\hat{\sigma}_+^{(1)} + e^{i(2\omega_a+\Sigma_2)t}\alpha^*_+\hat{\sigma}_+^{(2)}\\
  & \left.+ e^{i(2\omega_a-\Sigma_1)t}\alpha^*_-\hat{\sigma}_-^{(1)} + e^{i(2\omega_a-\Sigma_2)t}\alpha^*_-\hat{\sigma}_-^{(2)}\right] +h.c.,
\end{align}
where $\delta\Omega = (\Omega_1-\Omega_2)/2$ and $\Sigma_{1/2} = (3\Omega_{1/2}+\Omega_{2/1})/2$. If $\bar{g} > \delta\Omega$ we cannot ignore terms that oscillate at $\pm\delta\Omega$, as they do not average away over timescales of interest. All other terms are manifestly fast oscillating and will average away, except for the terms in the last line at frequencies $2\omega_a - \Sigma_{1/2}$. For these to be fast oscillating, we require that $2\omega_a - \Sigma_{1/2} \gg \bar{g}$. This simply implies that when designing the system, we must be careful to avoid the ``resonance'' conditions $2\omega_a = \Sigma_{1/2}$, so as not to activate unwanted qubit evolution.

Having dropped all fast oscillating terms the Hamiltonian bcomes
\begin{align}
  \nonumber\hat{H}' = \bar{g}\hat{a}^\dagger \left[e^{i\delta\Omega t}\left(\alpha_+\hat{\sigma}_+^{(1)} + \alpha_-\hat{\sigma}_-^{(2)}\right)\right. \\
  \left.+ e^{-i\delta\Omega t}\left(\alpha_+\hat{\sigma}_+^{(2)} + \alpha_-\hat{\sigma}_-^{(1)}\right)\right] +h.c.. \label{eqn:1C2Q}
\end{align}
Note that $\bar{g} > \delta\Omega$ is the opposite operating condition to the two cavity setup, where we require $\delta\Omega > \bar{g}$.

Assuming the cavity has a sufficiently broad linewidth, i.e.~$\delta\Omega \ll \kappa$, then the qubit processes in Eq.~\eqref{eqn:1C2Q} will interact with the environment through the cavity's nonzero density of states at $\pm\delta\Omega$, as depicted in Fig.~\ref{fig:2Q}c). If $2\delta\Omega$ is greater than the effective qubit linewidths, which are at least $4\alpha^2\bar{g}^2/\kappa$, then the processes at $+\delta\Omega$ and $-\delta\Omega$ will effectively see independent environments. In this case, adiabatically eliminating the cavity would result in a two-qubit master equation of the form of Eq.~\eqref{eqn:ME2}.

%%%%%%%%%%%%%%%%%%%%%%%%
\subsection{Two qubits and a transmission line}
\label{app:2QTL}

We consider two qubits coupled to a long waveguide or transmission
line (TL) described by the Hamiltonian
\begin{equation}
\hat{H}=\sum_{\alpha=1}^{2}\frac{\Omega_{\alpha}}{2}\hat{\sigma}_{z}^{(\alpha)}+\hat{H}_{TL}+\hat{H}_{I}.
\end{equation}
We consider the general case where the TL may be an engineered metamaterial, and hence is described by a general band structure describing propagating photonic modes (wavevector $k$, band index $\lambda$)
\begin{equation}
    \hat{H}_{TL}=\sum_{k\lambda}\omega_{k\lambda}\hat{b}_{k\lambda}^{\dagger}\hat{b}_{k\lambda},
\end{equation}
where we assume the dispersion relations satisfy  $\omega_{k,\lambda}=\omega_{-k,\lambda}$.
The qubit-TL couplings are described by
\begin{equation}
\hat{H}_{I}=\sum_{\alpha=1}^{2}g_{\alpha}(t)\hat{\sigma}_{x}^{(\alpha)}\hat{B}_{\alpha},
\end{equation}
where 
\begin{equation}
\hat{B}_{\alpha}=\sum_{k\lambda}\left(\frac{1}{\sqrt{L}}e^{ikx_{\alpha}}f_{k\lambda}\hat{b}_{k\lambda}^{\dagger}+\mathrm{h.c.}\right),
\end{equation}
with $x_{\alpha}$ denoting the position of the $\alpha$ qubit along the TL.  $f_{k\lambda}$ is a wavevector
and band dependent coupling coefficient which we assume to satisfy $f_{k\lambda}=f_{-k\lambda}^*$. 
To realize the synthetic squeezing dissipator, the qubit-TL
couplings $g_{\alpha}(t)$ are modulated at the sum of the qubit splitting
frequencies, and also have a static component: 
\begin{equation}
g_{\alpha}(t)=\overline{g}\left(\eta_{\alpha}+2\epsilon_{\alpha}\cos\left[\left(\Omega_{1}+\Omega_{2}\right)t\right]\right).
\label{eq:appa-modded-coupling}
\end{equation}
Note that in this description $\overline{g}$ has units $\mathrm{Hz}\sqrt{\mathrm{m}}$.

After moving into the interaction picture with respect to the system
Hamiltonian $\hat{H}_{S}=\sum_{\alpha=1}^{2}(\Omega_{\alpha}/2)\hat{\sigma}_{z}^{(\alpha)}$
and the TL Hamiltonian $\hat{H}_{TL}$, the system-TL interaction
can be written as
\begin{align}
\hat{H}_{I,\mathrm{int}}/\overline{g} & =\left[\eta_{1}e^{i\Omega_{1}t}+\epsilon_{1}e^{-i\Omega_{2}t}+\epsilon_{1}e^{i\left(2\Omega_{1}+\Omega_{2}\right)t}\right]\hat{\sigma}_{+}^{(1)}\hat{B}_{1}(t)\nonumber \\
 & +\,\left[\eta_{2}e^{i\Omega_{2}t}+\epsilon_{2}e^{-i\Omega_{1}t}+\epsilon_{2}e^{i\left(\Omega_{1}+2\Omega_{2}\right)t}\right]\hat{\sigma}_{+}^{(2)}\hat{B}_{2}(t)\nonumber \\
 & +\,\mathrm{h.c.}
\end{align}
which is of the form 
\begin{equation}
\hat{H}_{I,\mathrm{int}}=\sum_{\alpha,\omega}e^{-i\omega t}\hat{A}_{\alpha}(\omega)\hat{B}_{\alpha}(t)
\end{equation}
with $\alpha\in\{1,2\}$  indexing the qubits and $\omega\in\{\pm\Omega_{1},\pm\Omega_{2},\pm(2\Omega_{1}+\Omega_{2}),\pm(\Omega_{1}+2\Omega_{2})\}$
indexing all of the relevant frequencies. The $\hat{A}_{\alpha}(\omega)$
are e.g., $\hat{A}_{1}(\Omega_{1})=\overline{g}\eta_{1}\hat{\sigma}_{-}^{(1)}$ and $\hat{A}_{2}(\Omega_{1})=\overline{g}\epsilon_{1}\hat{\sigma}_{+}^{(2)}$.
With $\hat{H}_{I,\mathrm{int}}$ written in this form and following
the master equation derivation of \cite{Breuer:2002}, we
find the system density matrix master equation to be 
\begin{align}
    \frac{d}{dt}\hat{\rho}(t)=\sum_{\alpha,\beta}\sum_{\omega}\Gamma_{\alpha\beta}(\omega)\left(\hat{A}_{\beta}(\omega)\hat{\rho}(t)\hat{A}_{\alpha}^{\dagger}(\omega)\right. \nonumber\\
    \left. -\:\hat{A}_{\alpha}^{\dagger}(\omega)\hat{A}_{\beta}(\omega)\hat{\rho}(t)\right)+\mathrm{h.c.}
\end{align}
where we have introduced the bath correlation functions
\begin{equation}
\Gamma_{\alpha\beta}(\omega)\equiv\int_{0}^{\infty}dse^{i\omega s}\langle\hat{B}_{\alpha}^{\dagger}(s)\hat{B}_{\beta}(0)\rangle.
\end{equation}
Here $\langle\hat{B}_{\alpha}^{\dagger}(s)\hat{B}_{\beta}(0)\rangle=\mathrm{Tr}_{TL}\{\hat{B}_{\alpha}^{\dagger}(s)\hat{B}_{\beta}(0)\hat{\rho}_{TL}\}$
and we have assumed the TL state $\hat{\rho}_{TL}$ is stationary
so $\langle\hat{B}_{\alpha}^{\dagger}(t)\hat{B}_{\beta}(t-s)\rangle=\langle\hat{B}_{\alpha}^{\dagger}(s)\hat{B}_{\beta}(0)\rangle$. 

We next
decompose the bath correlation functions as
\begin{equation}
\Gamma_{\alpha\beta}(\omega)=\frac{1}{2}\gamma_{\alpha\beta}(\omega)+iS_{\alpha\beta}(\omega),
\end{equation}
for $S_{\alpha\beta}(\omega)=(\Gamma_{\alpha\beta}(\omega)-\Gamma_{\beta\alpha}^{*}(\omega))/2i$
and $\gamma_{\alpha\beta}(\omega)=\Gamma_{\alpha\beta}(\omega)+\Gamma_{\beta\alpha}^{*}(\omega)$.
This decomposition separates the master equation
into dissipative terms with correlation function coefficients $\gamma_{\alpha\beta}(\omega)$
and Hamiltonian terms with correlation function coefficients $S_{\alpha\beta}(\omega)$.
The master equation can then be recast into the form
\begin{equation}
\frac{d}{dt}\hat{\rho}(t)=-i[\hat{H}_{LS},\hat{\rho}(t)]+D(\hat{\rho}(t)),
\end{equation}
where the coherent evolution of the system due to the transmission
line is given by the ``Lamb shift'' Hamiltonian
\begin{equation}
\hat{H}_{LS}=\sum_{\omega}\sum_{\alpha,\beta}S_{\alpha\beta}(\omega)\hat{A}_{\alpha}^{\dagger}(\omega)\hat{A}_{\beta}(\omega),
\end{equation}
and the dissipation is given by
\begin{align}
    D(\hat{\rho}(t))=\sum_{\omega}\sum_{\alpha,\beta}\gamma_{\alpha\beta}(\omega)\left(\hat{A}_{\beta}(\omega)\hat{\rho}(t)\hat{A}_{\alpha}^{\dagger}(\omega)\right.\nonumber\\
    \left.-\:\frac{1}{2}\{\hat{A}_{\alpha}^{\dagger}(\omega)\hat{A}_{\beta}(\omega),\hat{\rho}(t)\}\right).
\end{align}
The Lamb shift Hamiltonian contains two kinds of terms: local terms that correspond to energy shifts of each qubit, and more interestingly, a coherent qubit-qubit ``pairing''
interaction of the form 
\begin{equation}
\hat{H}_{\mathrm{qb-qb}}=\Lambda\hat{\sigma}_{+}^{(1)}\hat{\sigma}_{+}^{(2)}+\mathrm{h.c.}
\end{equation}
 The local qubit energy shifts simply renormalize
the bare qubit frequencies, and hence do not play any significant role. 
In contrast, the pairing interaction can in principle disrupt our entanglement stabilization scheme; we discuss this more later in this section.  Before doing this, we first analyze the form of the TL induced dissipation on the qubits.

The dissipative terms include dissipative processes whose rates are
governed by the TL correlation function $\gamma_{\alpha\beta}(\omega)$
for all possible frequencies $\omega\in\{\pm\Omega_{1},\pm\Omega_{2},\pm(2\Omega_{1}+\Omega_{2}),\pm(\Omega_{1}+2\Omega_{2})\}$.
We let the TL be at zero temperature so that 
\begin{equation}
\gamma_{\alpha\beta}(\omega<0)=0.
\end{equation}
Thus all negative frequency processes (which correspond to emission of photons from the TL) vanish. While this is necessary to obtain the exact synthetic squeezing dissipation we desire, in Sec.~\ref{sec:therm} we investigate the impact of a thermal environment on entanglement stabilization. 

The dissipative processes at frequency $2\Omega_{1}+\Omega_{2}$
($\Omega_{1}+2\Omega_{2}$) involve both single qubit loss on qubit 1 (2),
due to the static couplings $\eta_{i}$, as well as single qubit excitation on qubit
2 (1), due to the modulated couplings $\epsilon_{i}$. These dissipative processes
destroy the entanglement we seek to generate. Therefore we must engineer
the TL to have a vanishing density of states at these frequencies.  This could be arranged through standard filtering or bandgap engineering techniques.  
When the TL has zero density of states at these frequencies, the correlation
functions become
\begin{equation}
\gamma_{\alpha\beta}(2\Omega_{1}+\Omega_{2})=\gamma_{\alpha\beta}(\Omega_{1}+2\Omega_{2})=0.
\end{equation}

We are thus left with the dissipative processes at $\Omega_{1}$ and
$\Omega_{2}$. The $\Omega_{1}$ processes take the form
\begin{align}
D_{\Omega_{1}}(\hat{\rho}(t))/\overline{g}^{2} & =\eta_{1}^{2}\gamma_{11}(\Omega_{1})\mathcal{D}[\hat{\sigma}_{-}^{(1)}]\hat{\rho}(t)\nonumber \\
 & +\,\eta_{1}\epsilon_{2}\gamma_{12}(\Omega_{1})\mathcal{D}[\hat{\sigma}_{+}^{(2)},\hat{\sigma}_{-}^{(1)}]\hat{\rho}(t)\nonumber \\
 & +\,\eta_{1}\epsilon_{2}\gamma_{21}(\Omega_{1})\mathcal{D}[\hat{\sigma}_{-}^{(1)},\hat{\sigma}_{+}^{(2)}]\hat{\rho}(t)\nonumber \\
 & +\,\epsilon_{2}^{2}\gamma_{22}(\Omega_{1})\mathcal{D}[\hat{\sigma}_{+}^{(2)}]\hat{\rho}(t),
\end{align}
where $\mathcal{D}[\hat{O}]\hat{\rho}$ is the usual Lindblad dissipator
and $\mathcal{D}[\hat{O},\hat{P}]\hat{\rho}=\hat{O}\hat{\rho}\hat{P}^{\dagger}-\{\hat{P}^{\dagger}\hat{O},\hat{\rho}\}/2$.
The $\Omega_{2}$ process is the same but with the indices $1\leftrightarrow2$.
Note that if the correlation functions $\gamma_{\alpha\beta}(\Omega_{1})=\gamma_{1}$
were all equal, this dissipator would reduce to the collective dissipator
\begin{equation}
D_{\Omega_{1}}(\hat{\rho}(t))=\overline{g}^{2}\gamma_{1}\mathcal{D}[\eta_{1}\hat{\sigma}_{-}^{(1)}+\epsilon_{2}\hat{\sigma}_{+}^{(2)}]\hat{\rho}(t).
\end{equation}

The above correlation functions are readily evaluated in general (in the
continuum limit $\frac{1}{L}\sum_{k}\to\int\frac{dk}{2\pi}$) as
\begin{equation}
\gamma_{\alpha\beta}(\omega)=\sum_{\lambda}\sum_{j}\frac{|f_{\lambda}(k_{j})|^{2}}{v_{g,\lambda}(k_{j})}e^{ik_{j}(x_{\beta}-x_{\alpha})},
\end{equation}
where the $\{k_{j}\}$ are all wavevectors for which $\omega_{\lambda}(k)=\omega$
and $v_{g,\lambda}(k)=|\partial\omega_{\lambda}(k)/\partial k|$ is
the group velocity at frequency $\omega_{\lambda}(k)$. Then for the
$\Omega_{1}$ processes, we have
\begin{equation}
\gamma_{\alpha\beta}(\Omega_{1})=\frac{2}{v_{g}(k_{1})}\cos\left(k_{1}(x_{\beta}-x_{\alpha})\right),
\end{equation}
where we have dropped the band index $\lambda$ and absorbed the coupling
coefficient $f_{\lambda}(k)$ into the group velocity.
Here $k_{1}$ is the wavevector for which $\omega_{\lambda}(k_{1})=\Omega_{1}$
for the correct band $\lambda$. We get a similar result for the $\Omega_{2}$
processes. 

Defining the qubit spacing
\begin{equation}
l\equiv x_{2}-x_{1},
\end{equation}
we can rewrite the $\Omega_{1}$ dissipation as
\begin{align}
    &D_{\Omega_{1}}(\hat{\rho}(t))  =\frac{2\overline{g}^{2}}{v_{g}(k_{1})}\cos(k_{1}l)\mathcal{D}[\eta_{1}\hat{\sigma}_{-}^{(1)}+\epsilon_{2}\hat{\sigma}_{+}^{(2)}]\hat{\rho}(t)\nonumber \\
     &+\:\frac{2\overline{g}^{2}}{v_{g}(k_{1})}\left(1-\cos(k_{1}l)\right)\left(\eta_{1}^{2}\mathcal{D}[\hat{\sigma}_{-}^{(1)}]+\epsilon_{2}^{2}\mathcal{D}[\hat{\sigma}_{+}^{(2)}]\right)\hat{\rho}(t).
    \label{eq:TLOmega1Dissipation}
\end{align}
In the ideal case where $k_{1}l=\pi n$ for $n\in\mathbb{Z}$, this all reduces to a
single dissipator with collective jump operator 
\begin{equation}
\hat{\mathcal{J}}_{1}=\eta_{1}\hat{\sigma}_{-}^{(1)}\pm\epsilon_{2}\hat{\sigma}_{+}^{(2)},
\end{equation}
where the sign is set by the parity of $n$ ($+$ for even $n$).
In the more general case where $k_{1}l\neq\pi n$, Eq.~(\ref{eq:TLOmega1Dissipation}) tells us the dissipation will both have the desired collective dissipator, but also unwanted uncorrelated single qubit dissipators (loss on qubit $1$, excitation on qubit $2$). 
For a small spacing error $k_1 l = \pi n +\delta$ where $\delta\ll1$, these additional dissipators are given to leading order in $\delta$ by
\begin{align}
    D_{\Omega_1,{\rm bad}}(\hat{\rho}(t)) = \delta^2 \frac{2\bar{g}^2}{v_g(k_1)}\left(\eta_{1}^{2}\mathcal{D}[\hat{\sigma}_{-}^{(1)}]+\epsilon_{2}^{2}\mathcal{D}[\hat{\sigma}_{+}^{(2)}]\right)\hat{\rho}(t).
\end{align}
As we discuss further below, our scheme can still perform well if these terms are non-zero but have a small amplitude (i.e.~the error in the qubit spacing is small).  

A similar analysis holds for the dissipative processes involving a frequency $\Omega_{2}$  in the TL.  The dissipation reduces to a single collective jump operator 
\begin{equation}
\hat{\mathcal{J}}_{2}=\epsilon_{1}\hat{\sigma}_{+}^{(1)}\pm\eta_{2}\hat{\sigma}_{-}^{(2)},
\end{equation}
when $k_{2}l=\pi m$ for $m\in\mathbb{Z}$, with the sign set by the
parity of $m$. 

Finally, if the qubit frequencies are chosen such that the characteristic wavevectors $k_1$, $k_2$ satisfy $k_{1}l=\pi n$ and $k_{2}l=\pi m$ for some spacing $l$ and integers $n$ and $m$, the final master equation is
\begin{equation}
\frac{d}{dt}\hat{\rho}(t)=-i[\hat{H}_{LS},\hat{\rho}]+\Gamma_{1}\mathcal{D}[\hat{\mathcal{J}}_{1}]\hat{\rho}+\Gamma_{2}\mathcal{D}[\hat{\mathcal{J}}_{2}]\hat{\rho}.
\end{equation}
Identifying $\eta_{1}=\alpha_{-}$, $\eta_{2}=\beta_{-}$, $\epsilon_{1}=\beta_{+}$,
and $\epsilon_{2}=\alpha_{+}$, we arrive at the desired master equation
Eq.~\eqref{eqn:ME2} \emph{except }with the additional Lamb shift
Hamiltonian. Note that if the qubit spacing is such that $k_{1}l=\pi n$
and $k_{2}l=\pi m$ for $m$, $n$ of opposite parity, we must flip the modulation phase for one of the qubit-waveguide couplings Eq.~\eqref{eq:appa-modded-coupling}, e.g., $\epsilon_2 \mapsto -\epsilon_2$. The signs in the dissipators $\hat{\mathcal{J}}_1$, $\hat{\mathcal{J}}_2$ must be the same to stabilize the desired entangled state.

We now consider the Lamb shift Hamiltonian $\hat{H}_{LS}$. It takes
the general form
\begin{equation}
\hat{H}_{LS}=\sum_{\alpha=1}^{2}\Omega_{\alpha,LS}\hat{\sigma}_{z}^{(\alpha)}+\left(\Lambda\hat{\sigma}_{+}^{(1)}\hat{\sigma}_{+}^{(2)}+\mathrm{h.c.}\right),
\end{equation}
where $\Omega_{\alpha,LS}$ are the inconsequential single qubit frequency shifts, and the pairing interaction amplitude is
\begin{align}
    \Lambda/\overline{g}^{2}=\eta_{1}\epsilon_{2}\left(S_{12}(\Omega_{1})+S_{21}(-\Omega_{1})\right)\nonumber\\
    +\:\epsilon_{1}\eta_{2}\left(S_{21}(\Omega_{2})+S_{12}(-\Omega_{2})\right).
\end{align}
We show below that $\Lambda$ is real, thus the eigenstates of the pairing interaction are paired states:
\begin{equation}
    \ket{\psi_\pm} = \frac{1}{\sqrt{2}}\left(\ket{00} \pm \ket{11}  \right).
\end{equation}
Recall that synthetic squeezing generates the paired state given by Eq.~\eqref{eqn:EntQ}, where the relative phase between $\ket{00}$ and $\ket{11}$ is $\pm1$. Thus, one of the eigenstates of the pairing Hamiltonian is always the $r\to\infty$ limit of the synthetic squeezing steady state, so that as $r$ is made larger the Hamiltonian will induce less disruption to our entanglement scheme. For any $r<\infty$ it will reduce the purity of the steady state, so we now evaluate $\Lambda$ to determine the magnitude of this impact.

Unlike the dissipation correlation functions $\gamma_{\alpha\beta}(\omega)$
which are readily evaluated with minimal specification of the TL band
structure, the $S_{\alpha\beta}(\omega)$ cannot be evaluated without
specifying the band structure of the TL or the $k$-dependent coefficients
$f_{k\lambda}$ of the system-TL coupling.
Nevertheless, we can make some general statements about $S_{\alpha\beta}(\omega)$, and thus about $\Lambda$, given the assumptions we have already made about the TL. The $S_{\alpha\beta}(\omega)$ are given by
\begin{equation}
    S_{\alpha\beta}(\omega) = \sum_\lambda \,{\rm pv}\! \int_{-k_c}^{k_c} \frac{dk}{2\pi} \frac{e^{ik(x_\beta - x_\alpha)}}{\omega - \omega_\lambda(k)} |f_\lambda(k)|^2,
\end{equation}
where pv denotes the principal value and $k_c$ is the band edge cutoff. At the outset, we had assumed reciprocity of the TL which implies $\omega_\lambda(k) = \omega_\lambda(-k)$ and $f_\lambda(k) = f_\lambda(-k)$. Thus the integral can be rewritten as
\begin{equation}
    S_{\alpha\beta}(\omega) = 2\sum_\lambda \,{\rm pv}\! \int_0^{k_c} \frac{dk}{2\pi}  \frac{\cos (kl_{\alpha\beta})}{\omega - \omega_\lambda(k)} |f_\lambda(k)|^2,
\end{equation}
where $l_{\alpha\beta}\equiv x_\beta - x_\alpha$. Therefore we see that $S_{\alpha\beta}(\omega)$ are real, thus so is $\Lambda$.

We can say little more about the correlation functions $S_{\alpha\alpha}(\omega)$ in general without more details of the TL, as these correlation functions are highly dependent on the band structure of the TL. 
Fortunately the pairing amplitude $\Lambda$ contains only $S_{\alpha\neq\beta}(\omega)$ factors which have the oscillating term $\cos(kl)$ in their integrals. If we let the qubit spacing be large, $k_{j}l \gg 1$, then the support of the integrals will be in a narrow region around the pole $\omega - \omega_\lambda(k)$. 
This implies a few things: first, the negative frequency correlation functions are negligible $|S_{\alpha\beta}(-\Omega_j)| \ll |S_{\beta\alpha}(+\Omega_j)|$ since they never encounter a pole. Second, the contributions to the positive frequency correlation functions by all bands other than that which contains $\Omega_j$ are similarly negligible. Finally, even in the band that contains $\Omega_j$ the integral is not particularly sensitive to either the band edge nor the dispersion relation since the rapid oscillation and $1/k$ behavior drastically reduce the integral's support away from the pole.

With the above general observations, we evaluate a rough estimate of $\Lambda$ for a typical TL. Suppose the TL has a low frequency band ($\lambda=1)$ for which $\omega(k_j) = \Omega_j$ (dropping the band index). For concreteness let's consider $S_{12}(\Omega_1)$. We linearize the dispersion relation about the pole $\omega(k) = \Omega_1$ as $\omega(k)\approx \omega(k_1) + v_g(k_1)(k-k_1)$ (for a typical TL far from the cutoff, this is a reasonable approximation). 
Then $S_{12}(\Omega_1)$ is
\begin{equation}
    S_{12}(\Omega_1) \approx \frac{2}{v_g(k_1)} \,{\rm pv}\! \int_0^{k_c} \frac{dk}{2\pi} \frac{\cos(kl)}{k_1 - k} |f(k)|^2.
\end{equation}
Finally, we assume that $f(k) \approx {\rm const}$ in the region around $k_1$ and absorb it into the group velocity. Since the integrand oscillates rapidly and quickly dies off away from $k=k_1$, we extend the integration bounds to $\pm\infty$, yielding
\begin{equation}
    S_{12}(\Omega_1) \approx \frac{1}{v_g(k_1)}\sin(k_1 l).
\end{equation}
A similar result holds for $S_{21}(\Omega_2)$. Therefore, we find the pairing amplitude $\Lambda$ to be
\begin{equation}
    \Lambda/\bar{g}^2 \approx \frac{\eta_1\epsilon_2}{v_g(k_1)}\sin(k_1 l) + \frac{\eta_2\epsilon_1}{v_g(k_2)}\sin(k_2 l),
\end{equation}
which vanishes for $k_jl=\pi n_j$ (integers $n_j$), exactly the conditions needed to realize synthetic squeezing.

In general, $\Lambda$ won't vanish exactly at the synthetic squeezing conditions. However, numerical evaluation of $\Lambda$ for various $\omega(k)$ (examples considered include $\omega=v_g|k|$, $\omega=(v_g/a)|\sin ka|$, and $\omega =  v_g(|k| - \alpha k^2)$) and for moderate qubit spacing $k_j l \sim 10 \pi$ shows that, for spacing near the synthetic squeezing conditions, typically
\begin{equation}
    \Lambda/\bar{g}^2 \approx \frac{\eta_1\epsilon_2}{v_g(k_1)}(\sin(k_1 l) +\delta_1) + \frac{\eta_2\epsilon_1}{v_g(k_2)}(\sin(k_2 l) + \delta_2)
\end{equation}
where usually $|\delta_j| \simeq 10^{-3}$. Thus $\Lambda$ may not exactly vanish at the synthetic squeezing conditions, but its value is typically quite suppressed. 

To get some sense for how this scheme performs in the face of qubit spacing imperfections, $l = l_{\rm ideal} + \delta l$, we consider a simple model system and analyze its performance numerically. To be concrete, consider two superconducting qubits with frequencies $\Omega_1 = 4$~GHz and $\Omega_2 = 6$~GHz and a transmission line with a linear dispersion 
\begin{equation}
    \omega(k) = v_g |k|,
\end{equation}
over the range 0 to 10~GHz, and with a (sharp) cutoff at 10~GHz (we require the transmission line to have zero density of states in a stop-band spanning at least $2\Omega_1 + \Omega_2 = 14$~GHz and $\Omega_1 + 2\Omega_2 = 16$~GHz). We choose to space the qubits by a distance
\begin{equation}
    l = 4\pi/k_1 + \delta l,
\end{equation}
or approximately two wavelengths at $\Omega_1$ (three wavelengths at $\Omega_2$). The ideal spacing, $\delta l=0$, satisfies the synthetic squeezing conditions $k_1 l = 4\pi$ and $k_2 l = 6\pi$. We let $\bar{g}^2/v_g \equiv 1$ and set $\eta_1=\eta_2=\cosh(r)$ and $\epsilon_1=\epsilon_2=\sinh(r)$. Then to a very good approximation
\begin{equation}
    \Lambda = \sinh(r)\cosh(r)\left( \sin(k_1 \delta l) + \sin(k_2 \delta l) \right).
\end{equation}
Thus the master equation is given by
\begin{align}
    &\dot{\hrho} = -i[\hat{H},\hrho] + \sum_j 2\cos(k_j \delta l)\mathcal{D}[\hat{\mathcal{J}}_j]\hrho \label{eq:TL-model-qme} \\
    &+ 4\sin^2\left(\frac{k_1 \delta 1}{2}\right)\left(\cosh^2(r)\mathcal{D}[\hat{\sigma}_{-}^{(1)}]+\sinh^2(r)\mathcal{D}[\hat{\sigma}_{+}^{(2)}]\right)\hat{\rho} \nonumber \\
    &+ 4\sin^2\left(\frac{k_2 \delta l}{2}\right)\left(\cosh^2(r)\mathcal{D}[\hat{\sigma}_{-}^{(2)}]+\sinh^2(r)\mathcal{D}[\hat{\sigma}_{+}^{(1)}]\right)\hat{\rho}, \nonumber
\end{align}
with $\hat{H} = \Lambda(\hat{\sigma}_-^{(1)}\hat{\sigma}_-^{(2)} + {\rm h.c.})$ and $\hat{\mathcal{J}}_{1/2} = \cosh(r)\hat{\sigma}_-^{(1/2)} + \sinh(r)\hat{\sigma}_+^{(2/1)}$. 

\begin{figure}[t]
  \includegraphics[width=0.99\columnwidth]{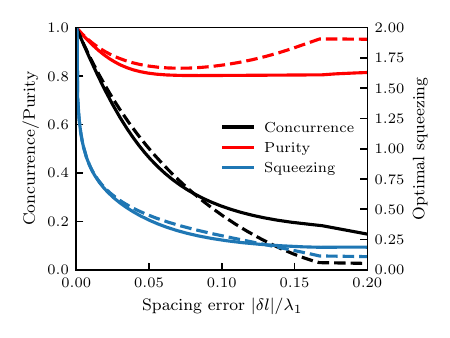}
  \caption{{\bf Performance of transmission line scheme with qubit spacing errors.} The purity (red) and optimal concurrence (black) of the steady state of Eq.~\eqref{eq:TL-model-qme} as a function of qubit spacing error $\delta l$ (measured as a fraction of the wavelength $\lambda_1$). The concurrence is optimized over the squeezing parameter $r$, and the optimal squeezing (blue) is plotted against the right axis. The optimal squeezing diverges as $\delta l\to 0$. The solid curves show the results of the full master equation and the dashed curves show the results of the master equation \emph{without} the Hamiltonian. The degradation of the entanglement is driven mostly by the single qubit dissipators.}
  \label{fig:TL-performance}
\end{figure}

We numerically compute the steady state of this master equation as a function of $\delta l$ and $r$. In general for $\delta l \neq 0$ the concurrence of the steady state is less than 1 but varies as a function of $r$. Thus, we numerically maximize the concurrence for a given $\delta l$ by optimizing over $r$. The results of this optimization are shown in Fig.~\ref{fig:TL-performance}. The concurrence, state purity, and optimal squeezing parameter are plotted against the spacing error $|\delta l|$ as a fraction of the wavelength $\lambda_1 = 2\pi/k_1$. In the plot, we compare the full master equation (solid curves) to the master equation wherein the Hamiltonian is excluded (dashed curves) to show that the single qubit dissipators cause the bulk of the performance degradation.

%%%%%%%%%%%
%%%%%%%%%%%

\section{Collective loss plus local qubit driving scheme: additional details}
\label{app:MS}

\subsection{Asymmetric waveguide couplings}

We consider the generalization of the scheme presented in Sec.~\ref{sec:collective-loss} to the general collective loss dissipator with arbitrary $\eta>0$.
In the main text we sought to provide the intuition for why this scheme works by exploiting the unitary equivalence between the family of the collective loss dissipator dark states Eq.~\eqref{eq:dark-family} and the family of target entangled states Eq.~\eqref{eq:entangled-family}.
Even when the loss dissipator is asymmetric, the local unitary equivalence still holds, but the local qubit rotations are no longer of equal angle but opposite sign. As finding the angles when $\eta\neq 1$ is nontrivial, we instead take a different approach. Given that the steady state we seek is simultaneously a dark state of the collective loss operator and a zero-energy eigenstate of the Hamiltonian, we can simply find the Hamiltonian parameters for which a given dark state of Eq.~\eqref{eq:collective-jump} is an eigenstate. We found that for the symmetric case the Hamiltonian is simply local Rabi drives on the qubits given by Eq.~\eqref{eq:coll-diss-H-og-frame}. We allow for a more general Hamiltonian
\begin{equation}
    \hat{H} = \frac{\Delta + \epsilon}{2}\hat{\sigma}_z^{(1)} - \frac{\Delta - \epsilon}{2}\hat{\sigma}_z^{(2)} + \Lambda^{(1)}\hat{\sigma}_x^{(1)} + \Lambda^{(2)}\hat{\sigma}_x^{(2)},
\end{equation}
wherein we allow for asymmetric Rabi drive amplitudes $\Lambda^{(j)}$ and asymmetric detunings parameterized by $\epsilon$.

The dark states of the collective of loss Eq.~\eqref{eq:collective-jump} can be expressed as
\begin{equation}
    \ket{\phi} = \alpha\ket{00} + \beta\left( \ket{01} - \eta\ket{10} \right),
\end{equation}
for $|\alpha|^2 + (1+\eta^2)|\beta|^2 = 1$. We take $\alpha,\beta>0$ real and positive without loss of generality. Enforcing that $\ket{\phi}$ is an eigenstate of $\hat{H}$, i.e.~$\hat{H}\ket{\phi} = E\ket{\phi}$, imposes the following constraints on the Hamiltonian parameters
\begin{align}
    &\Lambda^{(1)} = \eta \Lambda^{(2)}, \\
    &\epsilon = \frac{\Lambda^2}{\Delta}\left( 1 - \eta^2 \right),
    \label{eq:epsilon-def}
\end{align}
where in the latter we have defined $\Lambda\equiv\Lambda^{(2)}$. Imposing these conditions yields Eq.~\eqref{eq:asymmetricloss-H} as the final Hamiltonian. 

The unique steady state expressed in terms of the Hamiltonian parameters is given by
\begin{equation}
    \ket{\Phi} = \frac{1}{\sqrt{\Delta^2 +\Lambda^2(1+\eta^2)}}\left(\Delta\ket{00} + \Lambda\left( \ket{01} - \eta\ket{10} \right)\right).
\end{equation}
For symmetric loss, there is a frame in which this state is an entangled state of the form of Eq.~\eqref{eqn:EntQ}, and we now show this is also the case for asymmetric loss. As with the symmetric case, the correct frame will be defined by the eigenbasis of the Hamiltonian of Eq.~\eqref{eq:asymmetricloss-H}. This Hamiltonian is diagonalized by rotating about each qubit's $y$-axis through the angles
\begin{align}
    \theta_1 &= \arctan\left( \frac{2\eta\Lambda}{\Delta+\epsilon} \right), \\
    \theta_2 &= \arctan\left( \frac{-2\Lambda}{\Delta - \epsilon} \right),
\end{align}
with these angles lying on the principal branch of the arctangent. 

In the new frame, Eq.~\eqref{eq:asymmetricloss-H} transforms to
\begin{equation}
    \hat{H}'=\frac{\mu}{2}\left(\hat{\tau}_z^{(1)} - \hat{\tau}_z^{(2)}\right),
\end{equation}
for
\begin{equation}
    \mu = \sqrt{(\Delta - \epsilon)^2 +4\Lambda^2}.
\end{equation}
Making this transformation on the state $\ket{\Phi}$ yields our target state
\begin{equation}
    \ket{\Phi}' = \frac{1}{\sqrt{\beta_-^2 +\beta_+^2}}\left( \beta_-\ket{00}' - \beta_+\ket{11}' \right),
\end{equation}
where $\beta_->\beta_+$, and they are given by
\begin{align}
    \beta_- &= \mu + \Delta + \epsilon, \\
    \beta_+ &= \eta(\mu -\Delta +\epsilon).
\end{align}
$\ket{\Phi}'$ can be made to look explicitly like a squeezed state by introducing the squeezing parameter $r = {\rm artanh}\,\beta_+/\beta_-$. For the symmetric case ($\eta = 1$) this expression reduces to $r = (1/4)\ln(1+4\Lambda^2/\Delta^2)$ which is a rewriting of Eq.~\eqref{eq:HDriveParams}. In the general case, one cannot easily find the expressions for $\Delta$ and $\Lambda$ in terms of $r$, $\mu$, and $\eta$. Instead one must solve the following equations for $\Delta$ and $\Lambda$ in terms of the other parameters (recalling that $\epsilon$ is defined in terms of $\Delta$ and $\Lambda$, c.f. Eq.~\eqref{eq:epsilon-def})
\begin{align}
    \Delta&=-\frac{e^{2r}-1-\eta\left(1+e^{2r}\right)}{e^{2r}-1+\eta\left(1+e^{2r}\right)}\left(\mu+\epsilon\right), \label{eq:rabi-detuning-general} \\
    \Lambda&=\frac{1}{2}\sqrt{\mu^2 - (\Delta-\epsilon)^2}.
    \label{eq:rabi-drive-general}
\end{align}
Finally in the Rabi eigenbasis, the collective loss dissipator becomes
\begin{align}
    \hat{\mathcal{J}}' &= \hat{\mathcal{J}}'_Z + \hat{\mathcal{J}}'_1 + \hat{\mathcal{J}}'_2, \\
    \hat{\mathcal{J}}'_Z &= \frac{\eta\Lambda}{\mu}\left(\hat{\tau}_{z}^{(1)}-\hat{\tau}_{z}^{(2)}\right), \\
    \hat{\mathcal{J}}'_1 &= \frac{1}{2\mu}\left(\beta_{-}\hat{\tau}_{-}^{(1)}-\beta_{+}\hat{\tau}_{+}^{(2)}\right), \\
    \hat{\mathcal{J}}'_2 &= \frac{(\mu-\epsilon)^{2}-\Delta^{2}}{8\eta\mu\Lambda^{2}}\left(\beta_{-}\hat{\tau}_{-}^{(2)}-\beta_{+}\hat{\tau}_{+}^{(1)}\right).
\end{align}
This has precisely the same operator form as $\hat{\mathcal{J}}'$ given in the main text for the symmetric case. One difference that appears in the asymmetric case is that the coefficients of $\hat{\mathcal{J}}'_1$ and $\hat{\mathcal{J}}'_2$ differ. Only when $\eta = 1$ do the coefficients coincide. This has no effect on the steady state as the dissipation rates do not change the form of the steady state.

\subsection{Comparing dynamics against the ideal two-qubit squeezed dissipation master equation}
\label{app:MS2}

In Sec.~\ref{sec:collective-loss} we showed that with collective loss and local Rabi drives on two qubits, there is a frame in which the steady state is a two-qubit entangled state and that the master equation in this frame is given by
\begin{equation}
    \dot{\hrho}' = -i[\hat{H}',\hrho'] + 
        \Gamma 
            \mathcal{D}\left[\hat{\mathcal{J}}'_{1} + \hat{\mathcal{J}}'_{2} + \hat{\mathcal{J}}'_{Z} \right] \hrho',
    \label{eq:app-b-qme}
\end{equation}
with the Hamiltonian given by Eq.~\eqref{eq:xfrm-H-coll-loss} and the dissipator terms given by Eqs.~\eqref{eq:coll-loss-sq1}-\eqref{eq:coll-loss-dp}. We further showed that when the Hamiltonian energy $\mu\gg\Gamma$ (the collective loss rate), the three terms in the dissipator can be split into three separate dissipators, as cross-terms between them become fast-oscillating in the interaction frame defined by $\hat{H}'$. 

The resulting master equation, Eq.~\eqref{eq:coll-loss-qme2}, is nearly identical to the dynamics of the ideal two-qubit squeezing master equation, Eq.~\eqref{eqn:qTMS} (with $\gamma_1 = \gamma_2 = \Gamma$). 
Moving to the interaction frame of $\hat{H}'$ in Eq.~\eqref{eq:coll-loss-qme2} does not affect the dissipators, so that the only remaining difference between the two master equations is the addition of the correlated dephasing $\hat{\mathcal{J}}_Z'$ to Eq.~\eqref{eq:coll-loss-qme2}. We now show that this additional dissipation has virtually no qualitative or quantitative effect on the stabilization of the entangled two-qubit state, and thus that Eq.~\eqref{eq:app-b-qme} and Eq.~\eqref{eqn:qTMS} are essentially the same when $\mu\gg\Gamma$. 

\begin{figure}[t]
  \includegraphics[width=0.9\columnwidth]{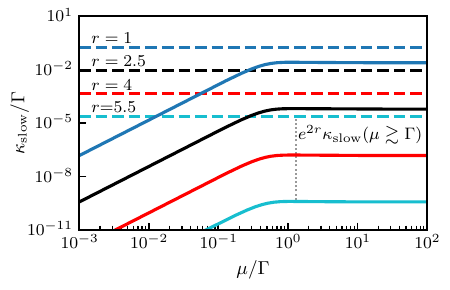}
  \caption{{\bf Comparison of dynamics between collective loss plus local driving and ideal two-qubit dissipative squeezing.} The dissipative gap $\kappa_{\rm slow}$ of Eq.~\eqref{eq:app-b-qme} as a function of the Hamiltonian energy $\mu$, computed numerically for various squeezing parameters $r$. The dashed lines show the dissipative gap of the ideal two-qubit squeezing master equation, Eq.~\eqref{eqn:qTMS} (with rates $\gamma_1=\gamma_2=\Gamma$). For small $\mu\ll\Gamma$ the dissipative gap vanishes as $\kappa_{\rm slow}\sim\mu^2$, and for large $\mu\gg\Gamma$ the gap saturates at a constant which is smaller than the ideal master equation gap by a factor of $e^{-2r}$. This further exponential suppression of the gap is due to the overall prefactors of $e^{-r}$ in the squeezing dissipators $\hat{\mathcal{J}}_{1/2}'$ (c.f. Eqs.~\eqref{eq:coll-loss-sq1} and \eqref{eq:coll-loss-sq2}).}
  \label{fig:coll-loss-diss-gap}
\end{figure}

As we discussed in Sec.~\ref{sec:SSE}, Eq.~\eqref{eqn:qTMS} has a dissipative gap $\kappa_{\rm slow}$ which is exponentially small in the squeezing parameter $r$, which is the limiting factor for relaxation to steady-state. The gap is thus the relevant metric for how closely Eq.~\eqref{eq:coll-loss-qme2} approximates the dynamics of Eq.~\eqref{eqn:qTMS}. To compare, we numerically calculate the dissipative gap of Eq.~\eqref{eq:app-b-qme} as a function of $\mu/\Gamma$ and compare it to the gap of Eq.~\eqref{eqn:qTMS}. The results of this are shown in Fig.~\ref{fig:coll-loss-diss-gap}, where the solid curves show the gap for various $r$ and the dashed lines show the gap of Eq.~\eqref{eqn:qTMS}. Recall that the gap closes at $\mu = 0$ (no Hamiltonian) because the collective loss dissipator has multiple steady states. 

A simple perturbative argument suggests that the gap should open as $\kappa_{\rm slow}\sim \mu^2$ which is precisely what is observed for $\mu\ll\Gamma$. When $\mu\gg\Gamma$ the gap saturates at a value $e^{-2r}$ times smaller than the ideal master equation gap. This factor comes from the prefactors of the jump operators $\hat{\mathcal{J}}'_{1/2}$ given in Eqs.~\eqref{eq:coll-loss-sq1} and \eqref{eq:coll-loss-sq2}. This extra suppression of the dissipative gap, and hence the state stabilization rate, is another important consideration for the practical implementation of this scheme, in addition to those discussed at the end of Sec.~\ref{sec:collective-loss}.

%%%%%%%%%%%
%%%%%%%%%%%

\section{Balanced Two-Qubit Dissipators}
\label{app:Balance}

For a setup described by Eq.~\eqref{eqn:qTMS}, only in the infinite squeezing limit, $r\rightarrow \infty$, is the steady-state a maximally entangled state. However, for our system we have the freedom to choose the modulation amplitudes, and in particular could choose $\al_{-} = \beta_{-} = \al_{+} = \beta_{+} = \sqrt{\bar{m}}$. In this case the master equation becomes
\begin{align}
  \dot\hrho_q = \frac{4\bar{m}\bar{g}^2}{\kk}\left(\mathcal{D}\left[\hat{\sigma}^{(1)}_- +\hat{\sigma}^{(2)}_+\right] + \mathcal{D}\left[\hat{\sigma}^{(1)}_+ + \hat{\sigma}^{(2)}_-\right]\right)\hrho_q. \label{eqn:ME2max}
\end{align}
As this is the $r\rightarrow \infty$ continuation of Eq.~\eqref{eqn:qTMS}, one might be tempted to believe that its steady-state is the maximally entangled Bell-state $\ket{\Phi^-} = (\ket{00} - \ket{11})/\sqrt{2}$, regardless of the coherent amplitude $\sqrt{\bar{m}}$, which now only controls the speed of the dissipative dynamics.

However, while the state $\ket{\Phi^-} $ is indeed a steady-state of the system, it is not the unique steady-state, and in fact any two-qubit state of the form
\begin{align}
 \hrho_{\rm SS} = \nu\ketbra{\Phi^-}{\Phi^-} + \frac{1}{4}(1-\nu)\hat{\iden}, \label{eqn:DSss}
\end{align}
is a steady-state of Eq.~\eqref{eqn:ME2max}, where $-1/3 \leq \nu \leq 1$ to ensure $\hrho_{\rm SS}$ is a valid density matrix. This one parameter family of states includes the maximally entangled state $\ket{\Phi^-}$ ($\nu =1$), but also the maximally mixed state ($\nu= 0$). As such, for the purposes of generating steady-state qubit entanglement, it is important to apply unbalanced drives to preserve the uniqueness of the steady-state, and avoid the degeneracy described above.

\end{document}